\documentclass[sigconf]{acmart}

\usepackage{subcaption}
\usepackage{bm}
\usepackage{algorithm}
\usepackage{algorithmic}
\usepackage{bbm}
\usepackage{makecell}
\usepackage{multirow}
\usepackage{balance}

\copyrightyear{2020} \acmYear{2020} \setcopyright{acmcopyright}\acmConference[SIGIR '20]{Proceedings of the 43rd International ACM SIGIR Conference on Research and Development in Information Retrieval}{July 25--30, 2020}{Virtual Event, China} \acmBooktitle{Proceedings of the 43rd International ACM SIGIR Conference on Research and Development in Information Retrieval (SIGIR '20), July 25--30, 2020, Virtual Event, China} \acmPrice{15.00} \acmDOI{10.1145/3397271.3401045} \acmISBN{978-1-4503-8016-4/20/07}

\begin{document}

\fancyhead{}
\renewcommand{\shortauthors}{Dai et al.}
\newcommand{\ED}{\Delta_e}
\newcommand\blfootnote[1]{%
	\begingroup
	\renewcommand\thefootnote{}\footnote{#1}%
	\addtocounter{footnote}{-1}%
	\endgroup
}
\def\cd{\,,\,}
\def\cuhk{The Chinese University of Hong Kong}

 \title{Convolutional Embedding for Edit Distance}

%

\author{Xinyan DAI}
\affiliation{
	\institution{\cuhk}
}
\email{xydai@cse.cuhk.edu.hk}

\author{Xiao Yan}
\affiliation{
	\institution{\cuhk}
}
\email{xyan@cse.cuhk.edu.hk}
\authornote{Corresponding author.}

\author{Kaiwen Zhou}
\affiliation{
	\institution{\cuhk}
}
\email{kwzhou@cse.cuhk.edu.hk}

\author{Yuxuan Wang}
\affiliation{
	\institution{\cuhk}
}
\email{yxwang7@cse.cuhk.edu.hk}

\author{Han Yang}
\affiliation{
	\institution{\cuhk}
}
\email{hyang@cse.cuhk.edu.hk}

\author{James Cheng}
\affiliation{
	\institution{\cuhk}
}
\email{jcheng@cse.cuhk.edu.hk}

\begin{abstract}
Edit-distance-based string similarity search has many applications such as spell correction, data de-duplication, and sequence alignment. However, computing edit distance is known to have high complexity, which makes string similarity search challenging for large datasets. In this paper, we propose a deep learning pipeline (called CNN-ED) that embeds edit distance into Euclidean distance for fast approximate similarity search. A convolutional neural network (CNN) is used to generate fixed-length vector embeddings for a dataset of strings and the loss function is a combination of the triplet loss and the approximation error. To justify our choice of using CNN instead of other structures (e.g., RNN) as the model, theoretical analysis is conducted to show that some basic operations in our CNN model preserve edit distance. Experimental results show that CNN-ED outperforms data-independent CGK embedding and RNN-based GRU embedding in terms of both accuracy and efficiency by a large margin. We also show that string similarity search can be significantly accelerated using CNN-based embeddings, sometimes by orders of magnitude.  
\end{abstract}



\maketitle

\section{Introduction}\label{sec:intro}

Given two strings $s_x$ and $s_y$, their edit distance $\ED(s_x, s_y)$ is the minimum number of edit operations (i.e., insertion, deletion and substitution) required to transform $s_x$ into $s_y$ (or $s_y$ into $s_x$). As a metric, edit distance is widely used to evaluate the similarity between strings. Edit-distance-based string similarity search has many important applications including spell corrections, data de-duplication, entity linking and sequence alignment~\cite{pivotal,  survey, empirical}.

The high computational complexity of edit distance is the main obstacle for string similarity search, especially for large datasets with long strings. For two strings with length $l$, computing their edit distance has $\mathcal{O}(l^2/\log(l))$ time complexity using the best algorithm known so far~\cite{faster-ed}. There are evidences that this complexity cannot be further improved~\cite{edit_quadratic}. Pruning-based solutions have been used to avoid unnecessary edit distance computation~\cite{passjoin,scalingup, edjoin,qchunk,adapt}. However, it is reported that pruning-based solutions are inefficient when a string and its most similar neighbor have a large edit distance~\cite{embedjoin}, which is common for datasets with long strings.          


Metric embedding has been shown to be successful in bypassing distances with high computational complexity (e.g., Wasserstein distance~\cite{wasserstein}). For edit distance, a metric embedding model can be defined by an embedding function $f(\cdot)$ and a distance measure $d(\cdot, \cdot)$ such that the distance in the embedding space approximates the true edit distance, i.e., $\ED(s_x, s_y) \!\approx\! d\left(f(s_x), f(s_y)\right)$. A small approximation error ($|\ED(s_x, s_y)-d\left(f(s_x), f(s_y)\right)|$) is crucial for metric embedding. For similarity search applications, we also want the embedding to preserve the order of edit distance. That is, for a triplet of strings, $s_x$, $s_y$ and $s_z$, with $\ED(s_x, s_y)<\ED(s_x, s_z)$, it should ensure that $d\left(f(s_x), f(s_y)\right)\!<\!d\left(f(s_x), f(s_z)\right)$. In this paper, we evaluate the accuracy of the embedding methods using both approximation error and order preserving ability.

Several methods have been proposed for edit distance embedding. Ostrovsky and Rabani embed edit distance into $\ell_1$ with a distortion\footnote{An embedding method is said to have a distortion of $\gamma$ if there exists a positive constant $\lambda$ that satisfies $\lambda \ED(s_x, s_y) \!\le\!d\left(f(s_x), f(s_y)\right)\!\le \!\gamma \lambda \ED(s_x, s_y)$, in which $\lambda$ is a scaling factor~\cite{wasserstein}.} of $2^{\mathcal{O}(\sqrt{\log l \log \log l})}$~\cite{l1embedding} but the algorithm is too complex for practical implementation. The CGK algorithm embeds edit distance into Hamming distance and the distortion is $\mathcal{O}(\ED)$~\cite{cgk}, in which $\ED$ is the true edit distance. CGK is simple to implement and shown to be effective when incorporated into a string similarity search pipeline. Both Ostrovsky and Rabani's method and CGK are data-independent while learning-based methods can provide better embedding by considering the structure of the underlying dataset. GRU~\cite{gru} trains a recurrent neural network (RNN) to embed edit distance into Euclidean distance. Although GRU outperforms CGK, its RNN structure makes training and inference inefficient. Moreover, its output vector (i.e., $f(s_x)$) has a high dimension, which results in complicated distance computation and high memory consumption. As our main baseline methods, we discussion CGK and GRU in more details in Section~\ref{sec:background}. 

To tackle the problems of GRU, we propose \textbf{CNN-ED}, which embeds edit distance into Euclidean distance using a convolutional neural network (CNN). The CNN structure allows more efficient training and inference than RNN, and we constrain the output vector to have a relatively short length (e.g., 128). The loss function is a weighted combination of the triplet loss and the approximation error, which enforces accurate edit distance approximation and preserves the order of edit distance at the same time. We also conducted theoretical analysis to justify our choice of CNN as the model structure, which shows that the operations in CNN preserve edit distance to some extent. In contrasts, similar analytical results are not known for RNN. As a result, we observed that for some datasets a randomly initialized CNN (without any training) already provides better embedding than CGK and fully trained GRU.

We conducted extensive experiments on 5 datasets with various cardinalities and string lengths. The results show that CNN-ED outperforms both CGK and GRU in approximation accuracy, computation efficiency, and memory consumption. The approximation error of CNN-ED can be only 50\% of GRU even if CNN-ED uses an output vector that is two orders of magnitude shorter than GRU. For training and inference, the speedup of CNN-ED over GRU is up to 30x and 200x, respectively. Using the embeddings for string similarity join, CNN-ED outperforms EmbedJoin~\cite{embedjoin}, a state-of-the-art method. For threshold based string similarity search, CNN-ED reaches a recall of 0.9 up to 200x faster compared with HSsearch~\cite{hstree}. Moreover, CNN-ED is shown to be robust to hyper-parameters such as output dimension and the number of layers.   

To summarize, we made three contributions in this paper. First, we propose a CNN-based pipeline for edit distance embedding, which outperforms existing methods by a large margin. Second, theoretical evidence is provided for using CNN as the model for edit distance embedding. Third, extensive experiments are conducted to validate the performance of the proposed method. 

The rest of the paper is organized as follows. Section~\ref{sec:background} introduces the background of string similarity search and two edit distance embedding algorithms, i.e., CGK and GRU. Section~\ref{sec:method} presents our CNN-based pipeline and conduct theoretical analysis to justify using CNN as the model. Section~\ref{sec:experiment} provides experimental results about the accuracy, efficiency, robustness and similarity search performance of the CNN embedding. The concluding remarks are given in Section~\ref{sec:conclusion}.

\section{Background and Related Work}\label{sec:background}

In this part, we introduce two string similarity search problems, and then discuss two existing edit distance embedding methods, i.e., CGK~\cite{cgk} and GRU~\cite{gru}.

\begin{algorithm}[]
	\caption{CGK Embedding}
	\label{alg:cgk}
	\begin{algorithmic}
		\STATE {\bfseries Input:} A string $s\!\in\!\mathcal{D}^l$ for some $ l\!\le\!L$, and a random matrix $\mathbf{R} \in \{0, 1\}^{3L \times |\mathcal{D}|}$
		\STATE {\bfseries Output:} An embedding sequence $y\!\in\!\{ \mathcal{D},  \perp \}^{3L |\mathcal{D}|} $
		\STATE Interpret $\mathbf{R}$ as $3L$ functions $\pi_1, \pi_2, \cdots, \pi_{3L}$, with $\pi_j(c_k) = \mathbf{R}_{jk}$, where $c_k$ denotes the $k^{\text{th}}$ character in $\mathcal{D}$  
		
		\STATE Initialize $i=0$, $y=\emptyset$ 
		\FOR{$j=1,2 \cdots, 3L$}
		\IF{$i\le l$ }
		\STATE $y=y\odot x[i]$ \quad	$\triangleright$ $\odot$ means concatenation
		\STATE $i = i + \pi_j(x[i])$						
	
		\ELSE 
		\STATE $y=y \  \odot \perp$			 
		\ \ \ \ \ \ \ \ \  $\triangleright$ pad with a special character $\perp$ 
		\ENDIF  
		\ENDFOR	
	\end{algorithmic}
\end{algorithm}

\subsection{String Similarity Search}
There are two well-known string similarity search problems, \textit{similarity join}~\cite{passjoin, edjoin, scalingup} and \textit{threshold search}~\cite{hstree}. For a dataset $\mathcal{S}=\{s_1,s_2,\cdots,s_n\}$ containing $n$ strings, similarity join finds all pairs $(s_i, s_j)$ of strings with $\ED(s_i,s_j)\!\le \!\tau$ and $i\!<\!j$, in which $\tau$ is a threshold for the edit distance between similar pairs. A number of methods~\cite{scalingup, passjoin, triejoin, duplicate, edjoin, massjoin, primitive-operator, embedjoin, minjoin} have been developed for similarity join but they are shown to be inefficient when the strings are long and $\tau$ is large. EmbedJoin~\cite{embedjoin} utilizes the CGK embedding~\cite{cgk} and  is currently the state-of-the-art method for similarity join on long strings. For a given query string $q$, threshold search~\cite{vgram, adapt, asymmetric, bedtree, pivotal, distributed_query} finds all strings $s\in \mathcal{S}$ that satisfies $\ED(q,s)\!\le \!\tau$. HSsearch~\cite{hstree} is one state-of-the-art method for threshold search, and outperforms Adapt~\cite{adapt}, QChunk~\cite{qchunk} and $B^{ed}$-tree~\cite{bedtree}. Similarity join is usually evaluated by the time it takes to find all similar pairs (called end-to-end time), while threshold search is evaluated by the average query processing time.

\subsection{CGK Embedding}

Algorithm~\ref{alg:cgk} describes the CGK algorithm~\cite{cgk}, which embeds edit distance into Hamming distance. It assumes that the longest string in the dataset $\mathcal{S}$ has a length of $L$ and the characters in the strings come from a known alphabet $\mathcal{D}$. $\mathbf{R}$ is a random binary matrix in which each entry is 0 or 1 with equal probability. $\perp\notin \mathcal{D}$ is a special character used for padding. Denote the CGK embeddings of two string $s_i$ and $s_j$ as $y_i$ and $y_j$, respectively. The following relation holds with high probability,
\begin{equation}\label{equ:CGK}
\ED(s_i, s_j)\le d_{\mathcal{H}}(y_i, y_j) \le \mathcal{O}(\ED^2(s_i, s_j)),
\end{equation}
in which $d_{\mathcal{H}}(y_i, y_j)=\sum_{k=1}^{3L}\mathbbm{1}[y_i(k)\neq y_j(k)]$ is the Hamming distance between $y_i$ and $y_j$.

\begin{figure}[t]	
	\centering
	\includegraphics[width=0.40\textwidth]{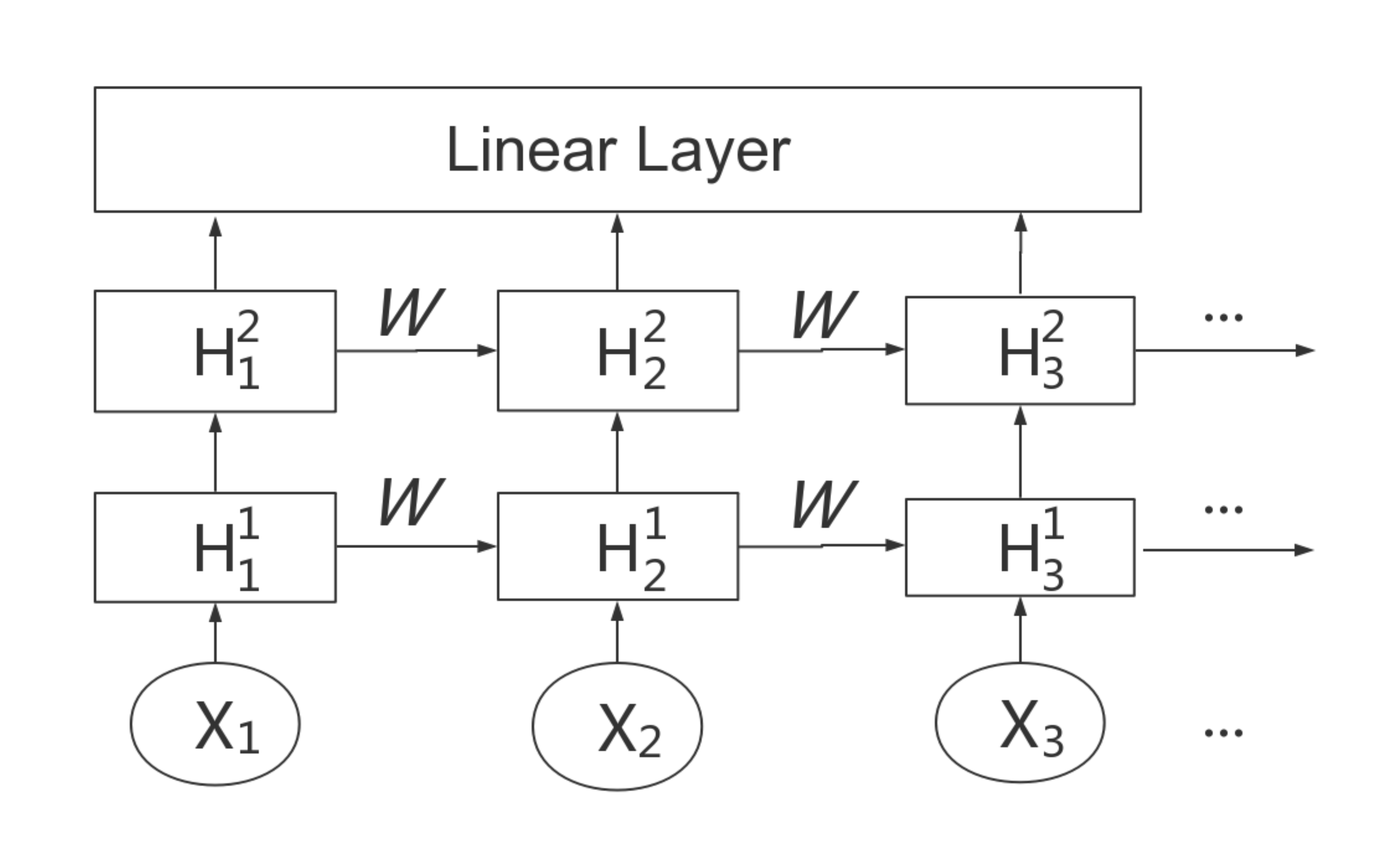}  
	\caption{The model architecture of GRU} 
	\label{fig:GRU}
\end{figure}

\begin{figure*}[h]	
	\centering 
	\includegraphics[width=0.85\textwidth]{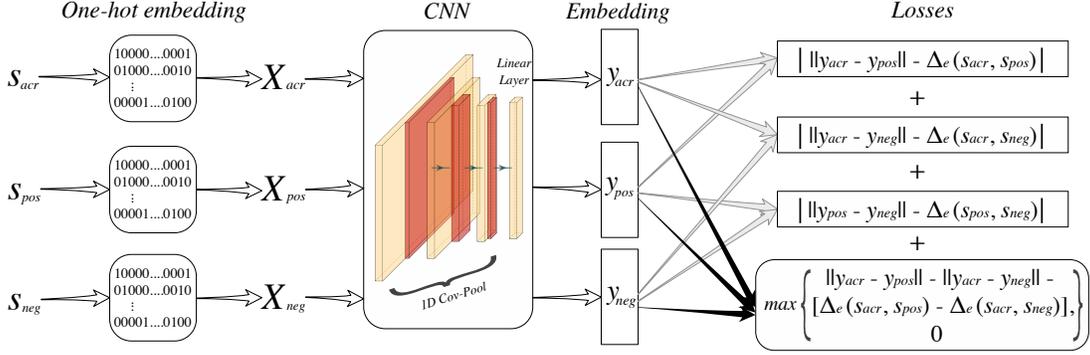}  
	\caption{The model architecture of CNN-ED} 
	\label{fig:arch}
\end{figure*}

\subsection{GRU Embedding}

RNN is used to embed edit distance into Euclidean distance in GRU~\cite{gru}. The network structure of GRU is shown in Figure~\ref{fig:GRU}, which consists of two layers of gated recurrent unit (GRU) and a linear layer. A string $s$ is first padded to a length of $L$ (the length of the longest string in the dataset)
and then each of its element is fed into the network per step. The outputs of the $L$ steps are concatenated as the final embedding. The embedding function of GRU can be expressed as follows
\begin{equation}
\begin{aligned}
h_i^1 = &GRU^1(s[i], h_{i-1}^1);\\
 h_i^2 = &GRU^2(h_i^1 \delta_i, h_{i-1}^2);\\  
f_{GRU}^{i} &= W h_i^2 + b;\\ 
f_{GRU} (s) & = [f_{GRU}^1, f_{GRU}^2, \cdots, f_{GRU}^L].
\end{aligned}
\end{equation}
As GRU uses the concatenation of the outputs, the embedding has a high dimension and takes up a large amount of memory. The network is trained with a three-phase procedure and a different loss function is used in each phase.

\section{CNN-ED}\label{sec:method}

We now present our CNN-based model for edit distance embedding. We first introduce the details of the learning pipeline, including input preparation, network structure, loss function and training method. Then we report an interesting phenomenon--a random CNN without training already matches or even outperforms GRU, which serves as a strong empirical evidence that CNN is suitable for edit distance embedding. Finally, we justify this phenomenon with theoretical analysis, which shows that operations in GNN preserves a bound on edit distances.

\subsection{The Learning Pipeline} 

We assume that there is a training set $\mathcal{S}\!=\!\{s_1,s_2,\cdots,s_n\}$ with $n$ strings. The strings (including training set, base dataset and possible queries) that we are going to apply our model on have a maximum length of $L$, and their characters come from a known alphabet $\mathcal{D}$ with size $|\mathcal{D}|$. $c_j$ denotes the $j^{\text{th}}$ character in $\mathcal{D}$. For two vectors $x$ and $y$, we use $\Vert x -y \Vert$ to denote their Euclidean distance.   

\medskip

\noindent\textbf{One-hot embedding as input.} For each training string $s_x$, we generate an one-hot embedding matrix $\bm{X}$ of size $|\mathcal{D}|\times L$ as the input for the model as follows,
\begin{equation}
\begin{aligned}
&\bm{X}=\big[\bm{X}_{1}, \cdots, \bm{X}_{|\mathcal{D}|}\big]^\intercal \text{with} \ \bm{X}_{j}\in \{0,1\}^L \ \text{for} \ 1\le j \le |\mathcal{D}|, \\
&\bm{X}_{j}[l]=\mathbbm{1}[s_x[l]=c_j] \ \ \text{for}  \ 1\le l \le L.
\end{aligned}
\end{equation}
For example, for $\mathcal{D}=$\{`A', `G', `C', `T'\} and $s_x=$``CATT'' and $L=4$, we have $\bm{X}=\left[
[0 1 0 0], 
[0 0 0 0],
[1 0 0 0],
[0 0 1 1]  \right]$. Intuitively, each row of $\bm{X}$ (e.g., $\bm{X}_{j}$) encodes a character (e.g., $c_j$) in $\mathcal{D}$, and if that character appears in certain position of $s_x$ (e.g., $l$), we mark the corresponding position in that row as 1 (e.g., $\bm{X}_{j}[l]=1$). In the example, the fourth row of $\bm{X}$ ($\bm{X}_{4}=[0 0 1 1]$) encodes the fourth character (i.e., `T'). $\bm{X}_{4}[3]=1$ and $\bm{X}_{4}[4]=1$ because `T' appears on the $3^{\text{rd}}$ and $4^{\text{th}}$ position of $s_x$. If string $s_{x}$ has a length $L'<L$, the last $L-L'$ columns of $\bm{X'}$ are filled with 0. In this way, we generate fixed-size input for the CNN. 

\medskip       

\noindent\textbf{Network structure.} The network structure of CNN-ED is shown in  Figure~\ref{fig:arch}, which starts with several one-dimensional convolution and pooling layers. The convolution is conducted on the rows of $\bm{X}$ and always uses a kernel size of 3. By default, there are 8 kernels for each convolutional layer and 10 convolutional layers. The last layer is a linear layer that maps the intermediate representations to a pre-specified output dimension of $d$ (128 by default). The one-dimensional convolution layers allow the same character in different positions to interact with each other, which corresponds to insertion and deletion in edit distance computation. As we will show in Section~\ref{subsec:theory}, max-pooling preserves a bound on edit distance. The linear layer allows the representation for different characters to interact with each other. Our network is typically small and the number of parameters is less than 45k for the DBLP dataset.

\medskip

\noindent\textbf{Loss function.} We use the following combination of triplet loss~\cite{tripletloss} and approximation error as the loss function
\newcommand{\loss}{\mathcal{L}}
\begin{equation*}
\loss(s_{acr}, s_{pos}, s_{neg})= \loss_t(s_{acr}, s_{pos}, s_{neg})+ \alpha \loss_p(s_{acr}, s_{pos}, s_{neg}),
\end{equation*}
in which $\loss_t$ is the triplet loss and $\loss_p$ is the approximation error. $(s_{acr}, s_{pos}, s_{neg})$ is a randomly sampled string triplet, in which $s_{acr}$ is the anchor string, $s_{pos}$ is the positive neighbor that has smaller edit distance to $s_{acr}$ than the negative neighbor $s_{neg}$. The weight $\alpha$ is usually set as 0.1. The triplet loss is defined as
\begin{equation*}
\loss_t(s_{acr}, s_{pos}, s_{neg})  \!=\! \max \left\lbrace0, 
\Vert y_{acr} \! - \! y_{pos} \Vert \!-\! \Vert y_{acr} \!-\! y_{neg} \Vert \!-\! \eta \right\rbrace,
\end{equation*}
in which $\eta=\ED(s_{acr}, s_{pos}) -  \ED(s_{acr}, s_{neg})$ is a margin that is specific for each triplet, and $y_{acr}$ is the embedding for $s_{acr}$. Intuitively, the triplet loss forces the distance gap in the embedding space ($ \Vert y_{acr}-y_{neg} \Vert-\Vert y_{acr}-y_{pos} \Vert$) to be larger than the edit distance gap ($\ED(s_{acr}, s_{neg})-\ED(s_{acr}, s_{pos})$), which helps to preserve the relative order of edit distance. The approximation error is defined as,    
\begin{equation*}
\loss_p(s_{acr}, s_{pos}, s_{neg})=w(s_{acr}, s_{pos})+w(s_{acr}, s_{neg})+w(s_{pos}, s_{neg}),
\end{equation*}                
in which $w(s_1, s_2)=\big| \Vert y_{s_1}-y_{s_2} \Vert - \ED(s_1, s_2)  \big|$ measures the difference between the Euclidean distance and edit distance for a string pair. Intuitively, the approximation error encourages the Euclidean distance to match the edit distance.

\medskip

\noindent\textbf{Training and sampling.} The network is trained using min-batch SGD and we sample 64 triplets for each min-batch. To obtain a triplet, a random string is sampled from the training set as $s_{acr}$. Then two of its top-$k$ neighbors ($k\!=\!100$ by default) are sampled, and the one having smaller edit distance with $s_{acr}$ is used as $s_{pos}$ while the other one is used as $s_{neg}$. For a training set with cardinality $n$, we call it an epoch when $n$ triplets are used in training.   

\medskip

\noindent\textbf{Using CNN embedding in similarity search.} The most straightforward application of the embedding is to use it to filter unnecessary edit distance computation. We demonstrate this application in Algorithm~\ref{alg:embeeding for search} for approximate threshold search. The idea is to use low-cost distance computation in the embedding space to avoid expensive edit distance computation. More sophisticated designs to better utilize the embedding are possible but is beyond the scope of this paper. For example, the embeddings can also be used to generate candidates for similarity search following the methodology of EmbedJoin, which builds multiple hash tables using CGK embedding and locality sensitive hashing (LSH)~\cite{lsh,crosslsh,lsh_survey}. To avoid computing all-pair distances in the embedding space, approximate Euclidean distance similarity methods such as vector quantization~\cite{pq,opq} and proximity graph~\cite{hnsw,nsg} can be used. Finally, it is possible to utilize multiple sets of embeddings trained with different initializations to provide diversity and improve the performance.

\begin{algorithm}[]
	\caption{Using Embedding for Approximate Threshold Search}
	\label{alg:embeeding for search}
	\begin{algorithmic}
		\STATE {\bfseries Input:} A query string $q$, a string dataset $\mathcal{S}\!=\!\{s_1,s_2,\cdots,s_n\}$, the embeddings of the strings $\mathcal{Y}\!=\!\{y_1,y_2,\cdots,y_n\}$, a model $f(\cdot)$, a threshold $K$ and a blow-up factor $\mu>1$  
		\STATE {\bfseries Output:} Strings with $\ED(q,s)\le K$
		\STATE Initialize the candidate set  $\mathcal{S}'=\emptyset$
		and result set as $\mathcal{C}=\emptyset$ 
		\STATE Compute the embedding of the query string $y_q=f(q)$
		\FOR{each emebdding $y_i$ in $\mathcal{Y}$}
		\IF{$\Vert y_q-y_i\Vert\le \mu\cdot K $}
		\STATE $\mathcal{S}'=\mathcal{S}'\cup s_i$
		\ENDIF 
		\ENDFOR
		\FOR{each string $s_i$ in $\mathcal{S}'$}  
		\IF{$\ED(q,s_i)\le K$}
		\STATE $\mathcal{C}=\mathcal{C}\cup s_i$ 
		\ENDIF  
		\ENDFOR	
	\end{algorithmic}
\end{algorithm}


\begin{figure}[!t]	
	\centering
	\includegraphics[width=0.23\textwidth]{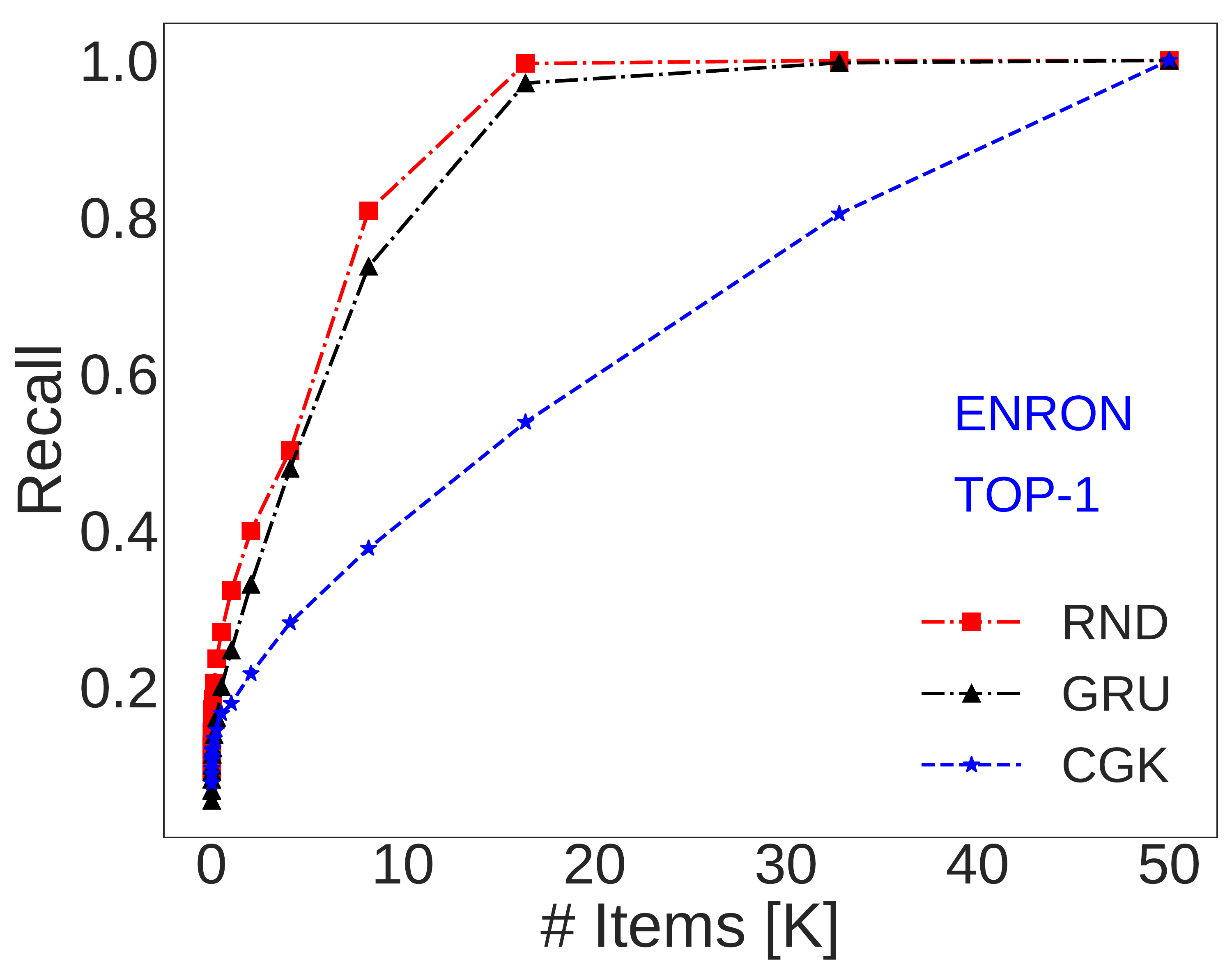}
	\includegraphics[width=0.23\textwidth]{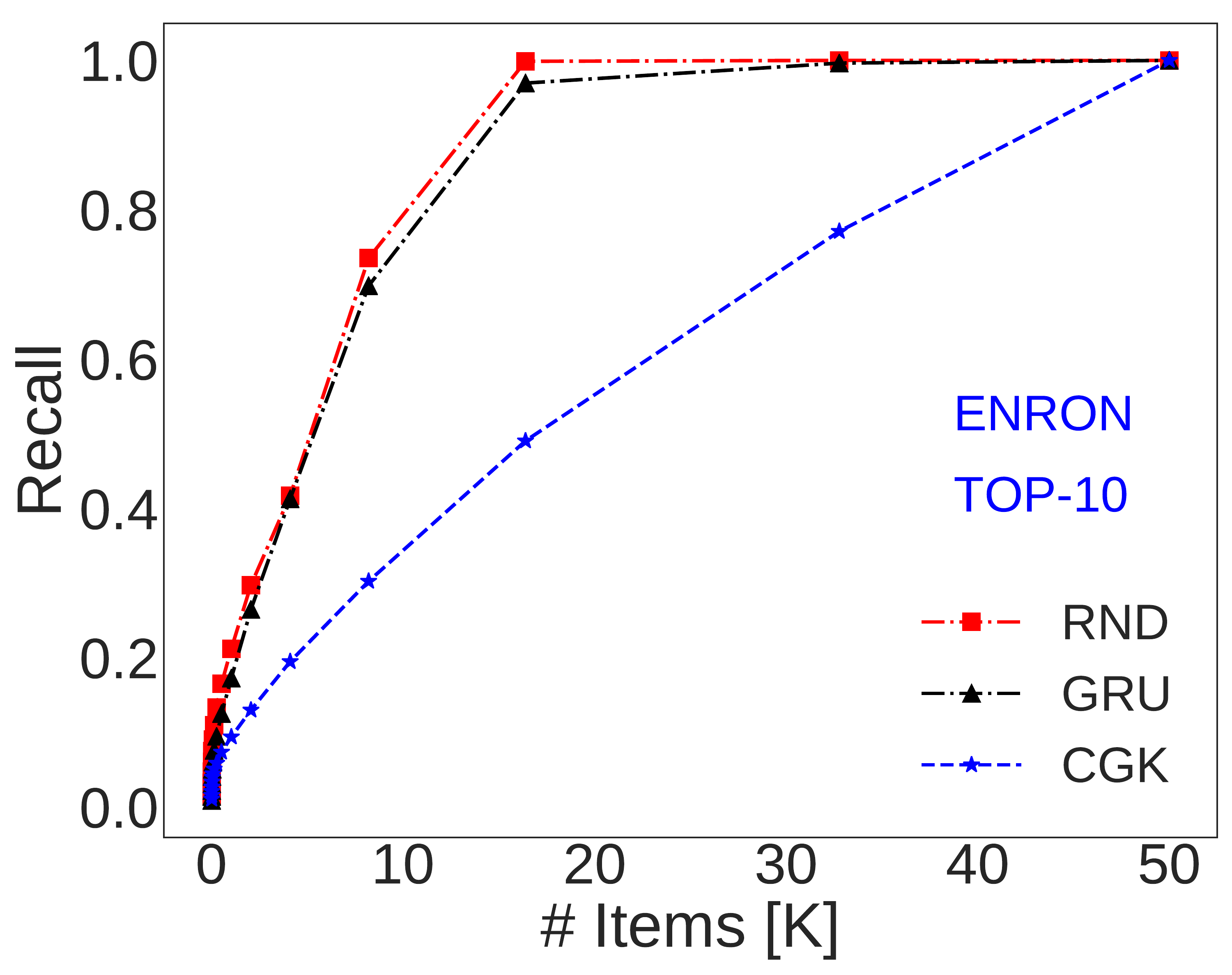} 
	\includegraphics[width=0.23\textwidth]{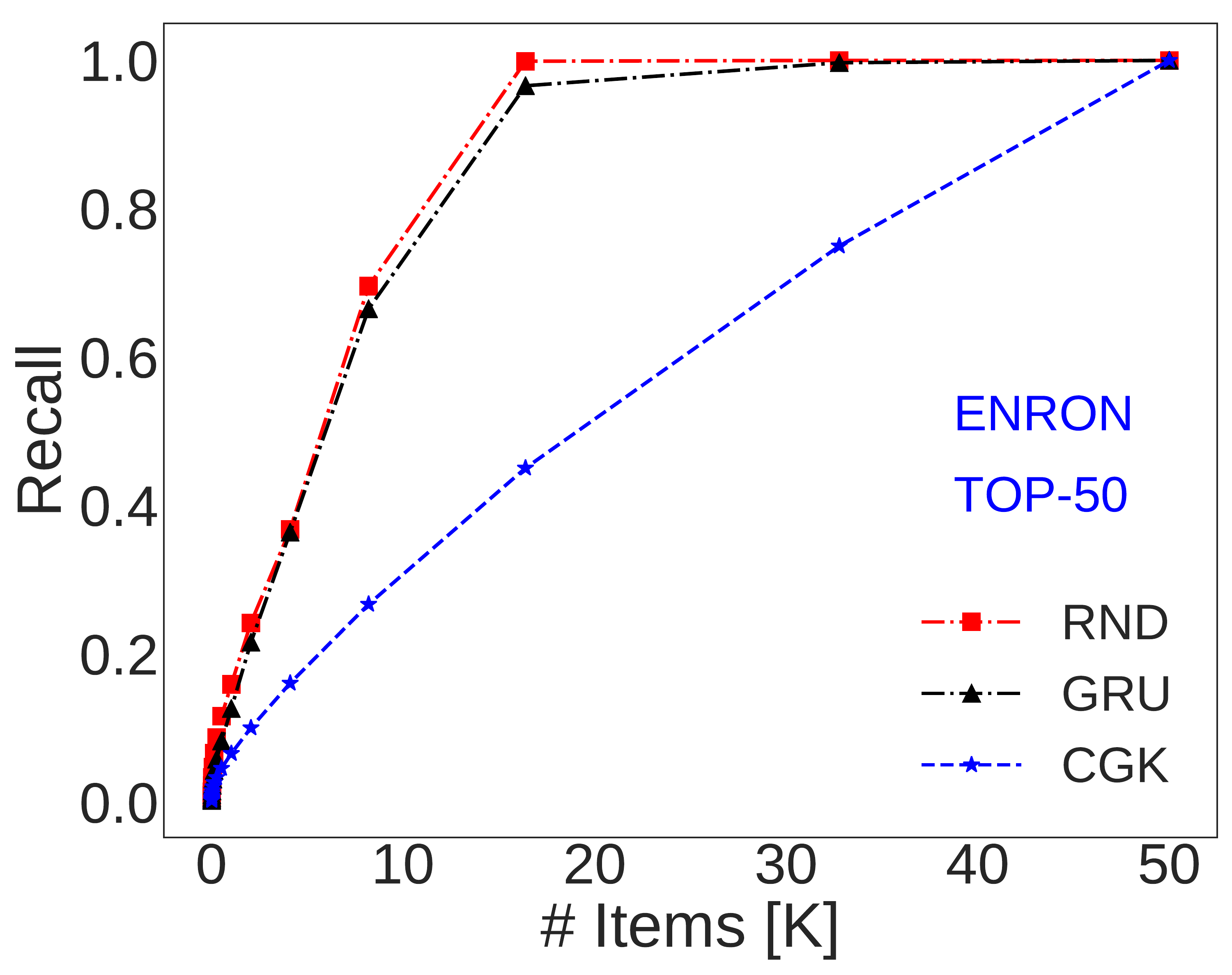} 
	\includegraphics[width=0.23\textwidth]{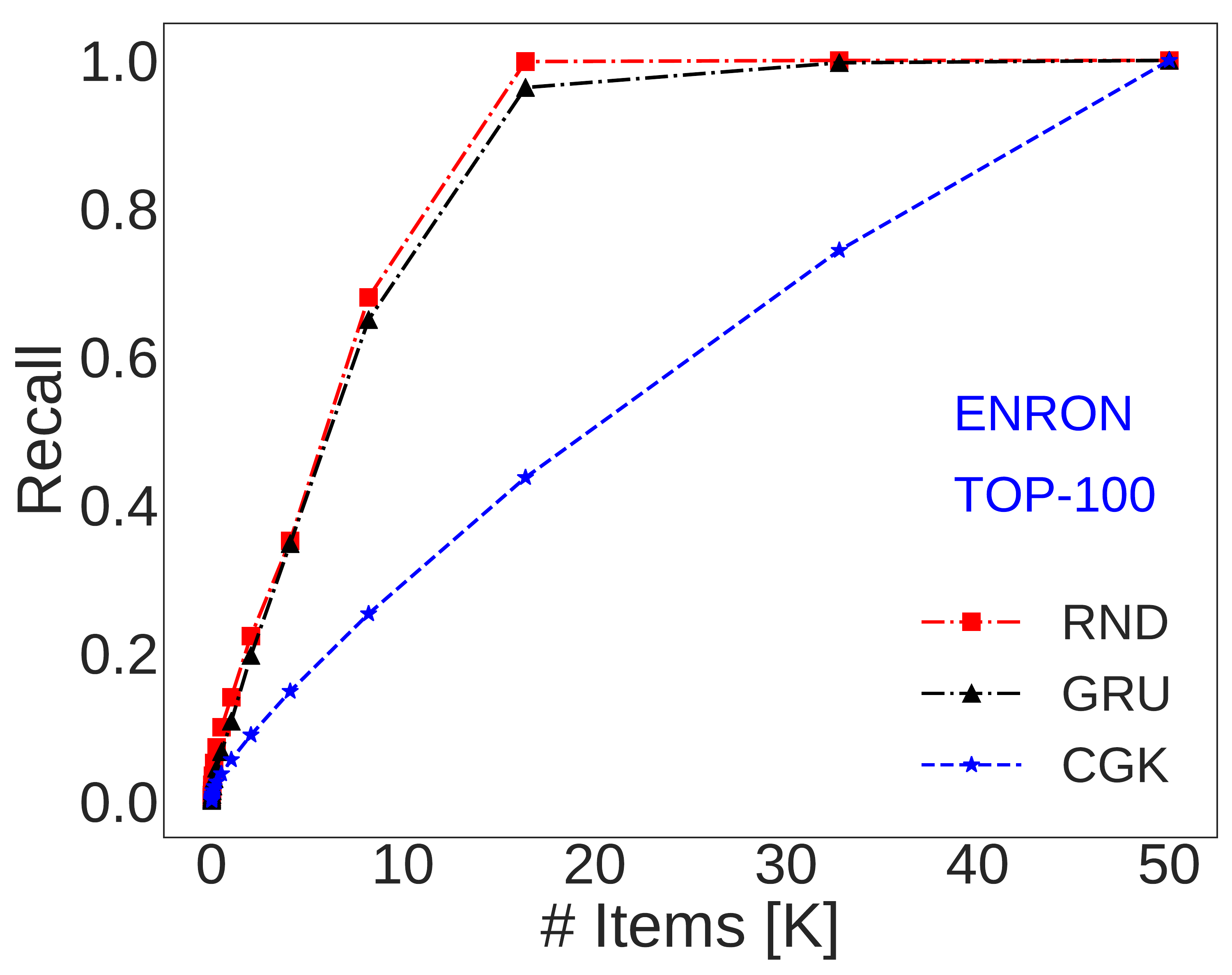} 
	\caption{Recall-item curve comparison for random CNN (denote as RND), CGK and GRU on the Enron dataset} 
	\label{fig:topk-enron random}
\end{figure}
\begin{figure}[!t]	
	\centering
	\includegraphics[width=0.23\textwidth]{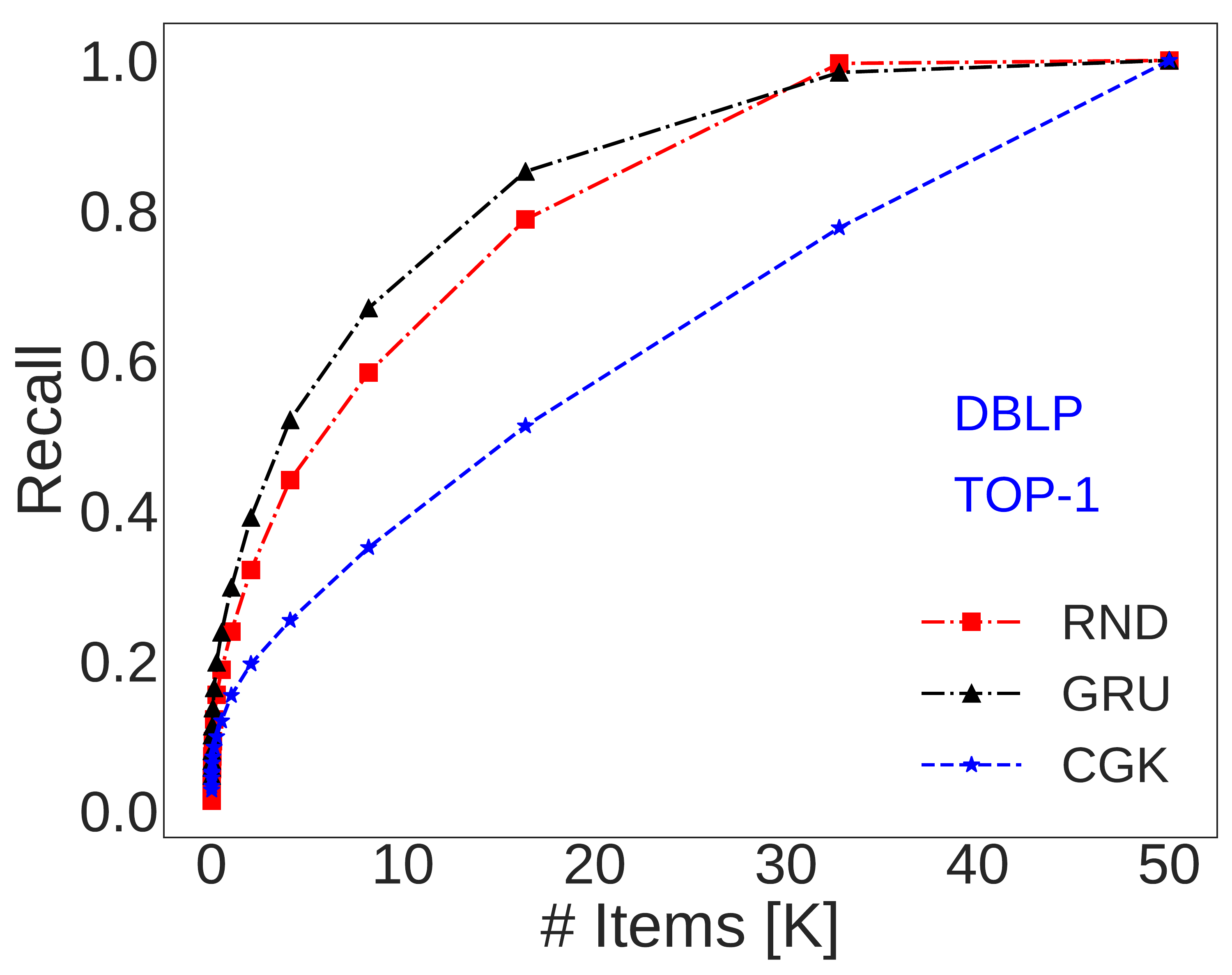}  
	\includegraphics[width=0.23\textwidth]{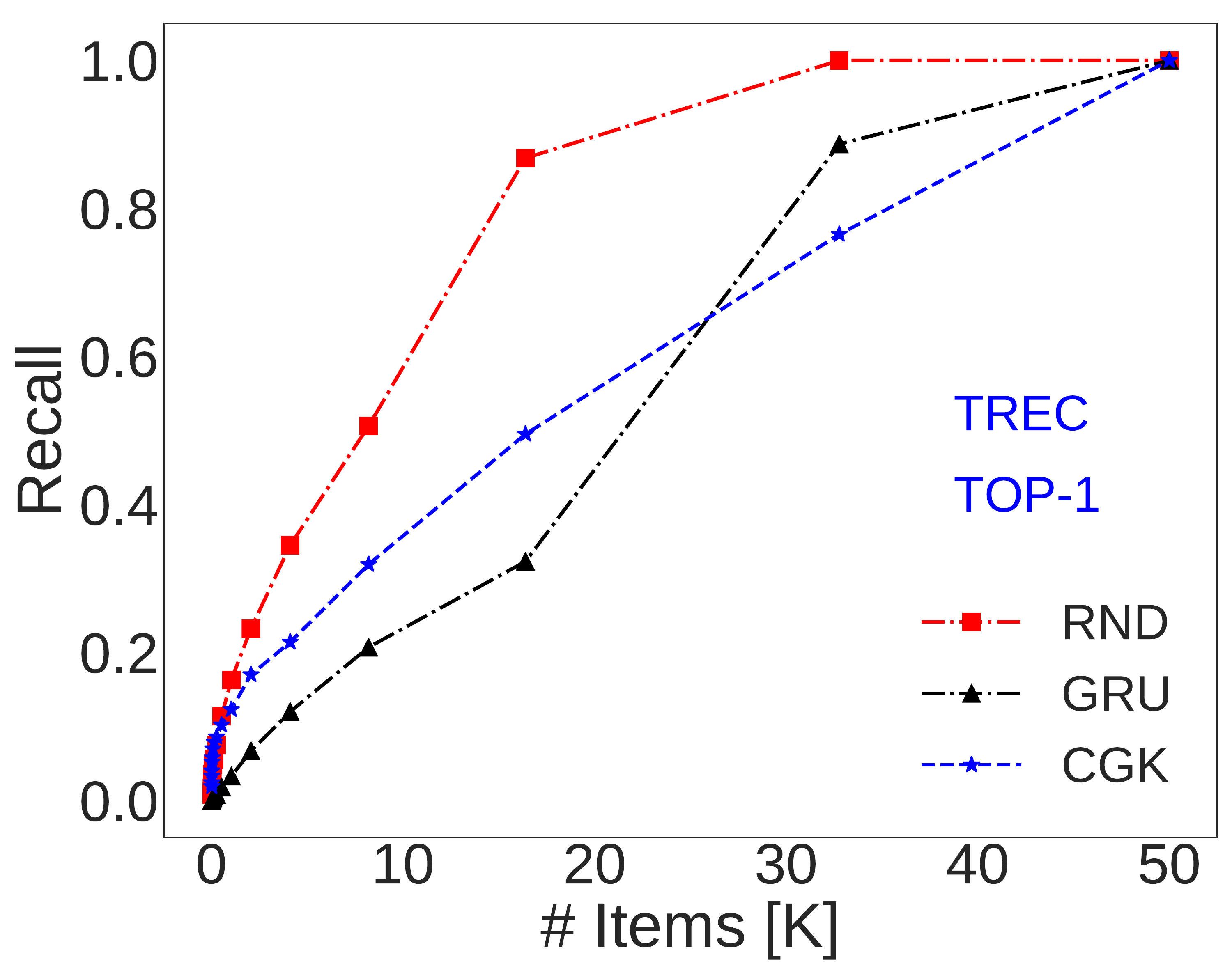} 
	\includegraphics[width=0.23\textwidth]{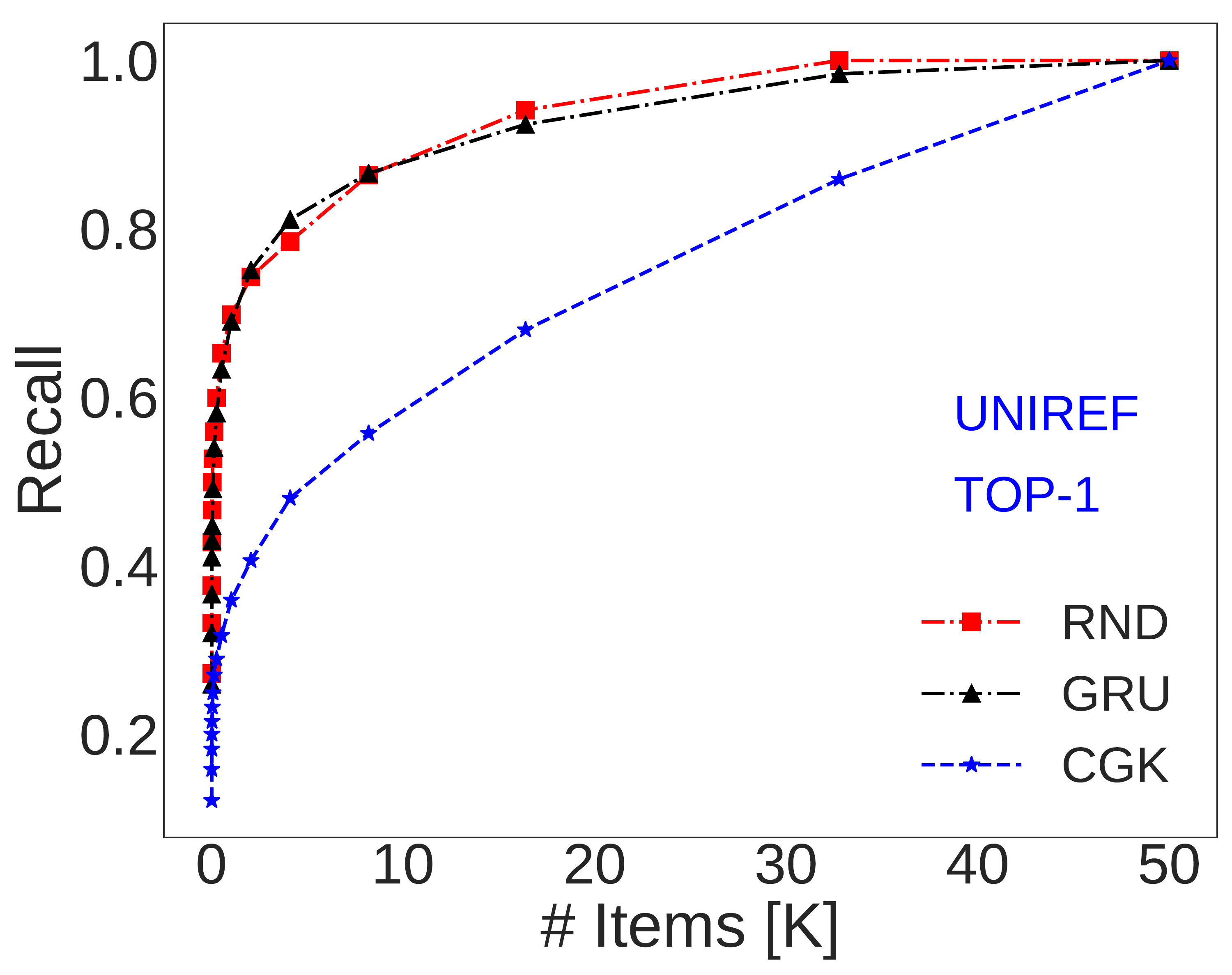}
	\includegraphics[width=0.23\textwidth]{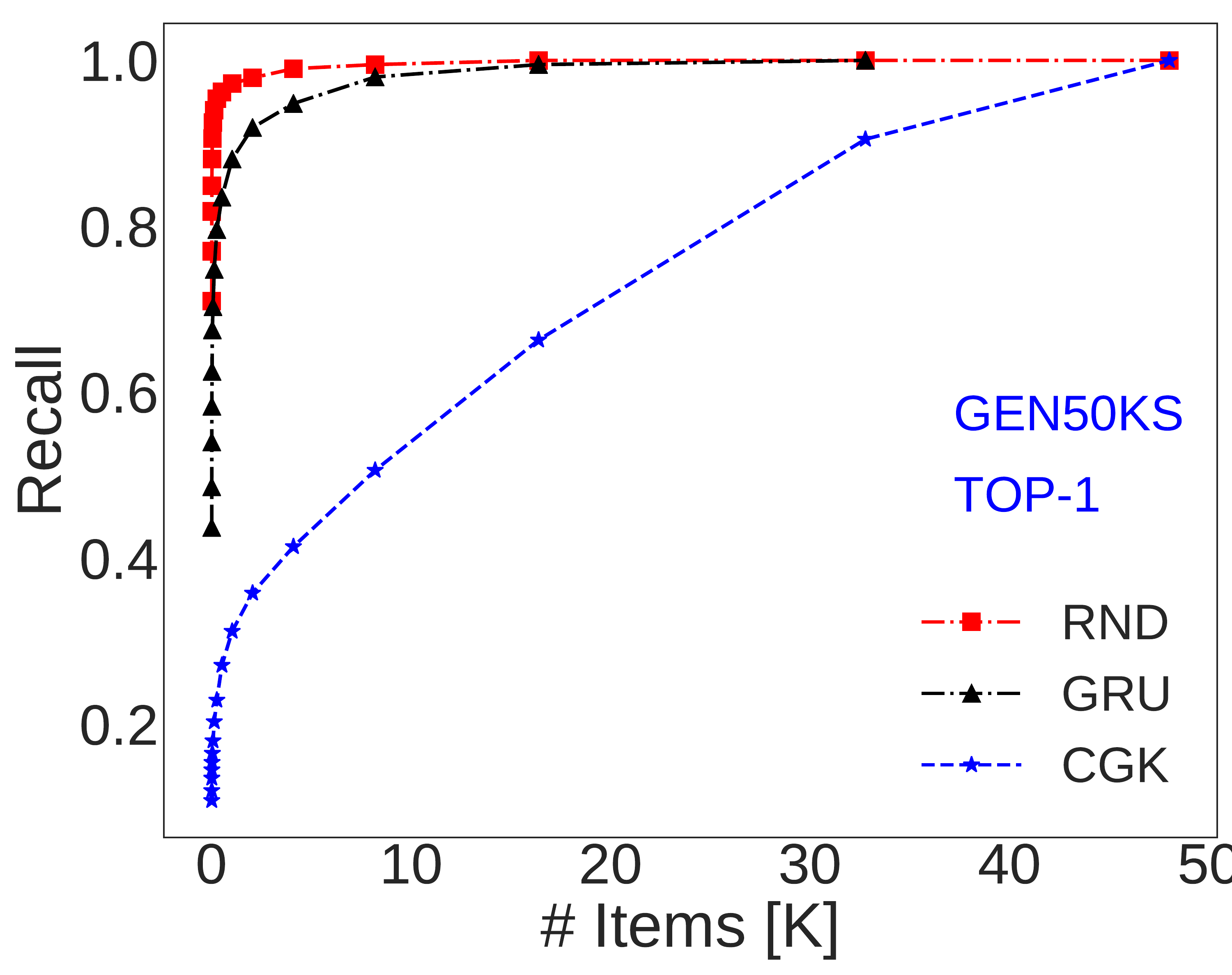}
	\caption{Recall-item curve comparison for random CNN (denote as RND), CGK and GRU on more datasets} 
	\label{fig:random-conv}
\end{figure}

\subsection{Why CNN is the Right Model?}\label{subsec:theory}

\noindent\textbf{Performance of random CNN.} In Figure~\ref{fig:topk-enron random} and Figure~\ref{fig:random-conv}, we compare the performance of CGK and GRU with a randomly initialized CNN, which has not been trained. The CNN contains 8 convolutional layers and uses max-pooling. The recall-item curve is defined in Section~\ref{sec:experiment} and higher recall means better performance. The statistics of the datasets can be found in Table~\ref{tab:statistic}. The results show that a random CNN already outperforms CGK on all datasets and for different value of $k$. The random CNN also outperforms fully trained GRU on Trec and Gen50ks, and is comparable to GRU on Uniref. Although random CNN does not perform as good as GRU on DBLP, the performance gap is not large. On the Enron dataset, random GNN slightly outperforms GRU for different values of $k$.

This phenomenon suggests that the CNN structure may have some properties that suit edit distance embedding. This is against common sense as strings are sequences and RNN should be good at handling sequences. To better understand this phenomenon, we analyze how the operations in our CNN model affects edit distance approximation. Basically, the results show that \textbf{one-hot embedding and max-pooling preserve bounds on edit distance}.


\newtheorem{one-hot}{Theorem}
\newtheorem{max-pool}[one-hot]{Theorem}

\begin{one-hot}[One-Hot Deviation Bound]
	\label{one-hot}
	Given two strings $s_x \in \{\mathcal{D}\}^M, s_y \in \{\mathcal{D}\}^N$ and their corresponding one-hot embeddings $\bm{X} = \big[\bm{X}_{1} ,\cdots ,\bm{X}_{|\mathcal{D}|}\big]^\intercal$ and $\bm{Y} = \big[\bm{Y}_{1} ,\cdots ,\bm{Y}_{|\mathcal{D}| }\big]^\intercal$, defining the binary edit distance as $\overline{\Delta}_e(s_x, s_y) \triangleq \sum_{i=1}^{|\mathcal{D}|}{\Delta_e\big(\bm{X}_{i}, \bm{Y}_{i}\big)}$, we have
	\[
	\begin{aligned}
	&|\mathcal{D}|\Delta_e(s_x, s_y) - (|\mathcal{D}| - 1) (M + N)\\\leq{}& \overline{\Delta}_e(s_x, s_y) \leq |\mathcal{D}| \Delta_e(s_x, s_y).	
	\end{aligned}
	\]
	\begin{proof}
		For the upper bound, note that by modifying the operations in the \textit{shortest edit sequence} \footnote{The edit sequence between two strings is a sequence of operations that transfer one string to the other one.}\footnote{The shortest edit sequence is one of the edit sequences with minimum length, i.e. the edit distance.} of changing $s_x$ into $s_y$ to binary operations, we can use this sequence to transform $\bm{X}_{i}$ into $\bm{Y}_{i}$, for any $i \in [|\mathcal{D}|]$. Since a substitution in the original sequence may be modified into `$0\rightarrow 0$', which is not needed, it satisfies that 
		\[
		\Delta_e\big(\bm{X}_{i}, \bm{Y}_{i}\big) \leq \Delta_e(s_x, s_y).
		\]
		Summing this bound for $i = 1, \ldots, |\mathcal{D}|$, we obtain the upper bound.
		
		For the lower bound, letting $s_x^{c_i}$ be the string of replacing the character in $s_x$ that is not $c_i$ with a special character $\perp \notin \mathcal{D}$, where $c_i$ is the $i^{\text{th}}$ character in the alphabet, we can conclude that 
		\[
		\begin{aligned}
		\Delta_e(s_x, s_x^{c_i}) &= M - |c_i|_{s_x},\\
		\Delta_e(s_x^{c_i}, s_y^{c_i}) &= \Delta_e\big(\bm{X}_{i}, \bm{Y}_{i}\big),
		\end{aligned}
		\]
		where $|c_i|_{s_x}$ is the number of character $c_i$ in $s_x$.
		
		Using the triangle inequality of edit distance, for any $i \in [|\mathcal{D}|]$, we have
		\[
		\begin{aligned}
		\Delta_e(s_x, s_y) &\leq \Delta_e(s_x, s_x^{c_i}) + \Delta_e(s_x^{c_i}, s_y) \\
		&\leq \Delta_e(s_x, s_x^{c_i}) + \Delta_e(s_x^{c_i}, s_y^{c_i}) + \Delta_e(s_y^{c_i}, s_y) \\
		&= M + N - |c_i|_{s_x} - |c_i|_{s_y} + \Delta_e\big(\bm{X}_{i}, \bm{Y}_{i}\big).
		\end{aligned}
		\]
		Summing this inequality for $i = 1, \ldots, |\mathcal{D}|$ and using that  
		\[
		\sum_{i=1}^{|\mathcal{D}|} {|c_i|_{s_x}} = M, \sum_{i=1}^{|\mathcal{D}|} {|c_i|_{s_y}} = N,
		\]
		we obtain
		\[
				|\mathcal{D}|\Delta_e(s_x, s_y) \leq |\mathcal{D}|(M + N) - M - N+ \overline{\Delta}_e(s_x, s_y).
		\]
		
		Re-arranging this inequality completes the proof.
		
	\end{proof}
\end{one-hot}

Note that the bound in Theorem~\ref{one-hot} can be tightened by choosing $\mathcal{D}$ as $\textup{supp}(s_x) \cup \textup{supp}(s_y)$. Theorem~\ref{one-hot} essentially shows that a bound on the true edit distance $\ED(s_x, s_y)$ can be constructed by the sum of the edit distances of $|\mathcal{D}|$ binary sequences. These binary sequences are exactly the rows of the one-hot embedding matrices $\bm{X}$ and $\bm{Y}$. This justifies our choice of using one-hot embedding as the input for the network.

\begin{max-pool}[Max-Pooling Deviation Bound]
	\label{max-pool}
	Given two binary vectors $x \in \{0,1\}^{M}, y \in \{0,1\}^N$ and a max-pooling operation $P(\cdot)$ on $x, y$ with stride $K$ and size $K$, assuming that $M$ and $N$ are divisible by $K$, the following holds:
	\[
	\begin{aligned}
	&\max{\left\{\begin{gathered}
		\Delta_e (x, y) - \frac{K-1}{K}(M + N),\\
		\frac{1}{K}\Delta_e (x, y) - \frac{K - 1}{K} (|1|_{P(x)} + |1|_{P(y)})
		\end{gathered}\right\}} \\
	\leq{}& \Delta_e(P(x), P(y)) \leq  \Delta_e (x, y) + \frac{K-1}{K}(M + N).
	\end{aligned}
	\]
	\begin{proof}
		Using the triangle inequality of edit distance, we have
		\[
		\begin{aligned}
		\Delta_e (x, y) &\leq \Delta_e(x, P(x)) + \Delta_e(P(x), y) \\
		&\leq \Delta_e(x, P(x)) + \Delta_e(P(x), P(y)) + \Delta_e (P(y), y) \\
		& = \frac{K-1}{K}(M + N) + \Delta_e(P(x), P(y)).
		\end{aligned}
		\]
		
		Applying this inequality again for $\Delta_e(P(x), P(y))$, we obtain
		\begin{equation}
		\label{P1}
		|\Delta_e (x, y) - \Delta_e(P(x), P(y))| \leq \frac{K-1}{K}(M + N).
		\end{equation}
		
		Denote $A(x)$ as the string of replicating each bit of $P(x)$ $K$ times. For the substitutions, insertions and deletions in the edit sequence of $\Delta_e(P(x), P(y))$, we can repeat these operations for the corresponding replicated bits in $A(x)$, which transform $A(x)$ into $A(y)$.
		Thus, we conclude that $\Delta_e(A(x), A(y)) \leq K\Delta_e(P(x), P(y))$. 
		
		Using triangle inequality, it satisfies that 
		\[
		\begin{aligned}
		\Delta_e (x, y) &\leq \Delta_e(x, A(x)) + \Delta_e(A(x), A(y)) + \Delta_e (A(y), y) \\
		&\leq \Delta_e(x, A(x)) + K\Delta_e(P(x), P(y)) + \Delta_e (A(y), y).
		\end{aligned}
		\]
		
		For $\Delta_e(x, A(x))$, if a bit is $0$ in $P(x)$, its corresponding window in $A(x)$ and $x$ must be all $0$; if the bit is $1$, the number of different bits in the corresponding window of $A(x)$ and $x$ is upper-bounded by $K - 1$, which implies that $\Delta_e(x, A(x)) \leq (K - 1) |1|_{P(x)}$, where $|1|_{P(x)}$ denotes the number of $1$ in $P(x)$. Thus,
		\[
		\Delta_e (x, y) \leq (K - 1) (|1|_{P(x)} + |1|_{P(y)}) + K\Delta_e(P(x), P(y)).
		\]
		
		Rearranging this lower bound and the bound \eqref{P1} complete the proof. 
	\end{proof} 
\end{max-pool}

Theorem~\ref{max-pool} shows that max-pooling preserves a bound on the edit distance of binary vectors. Combining with Theorem~\ref{one-hot}, it also shows that max-pooling preserves a bound on the true edit distance $\ED(s_x, s_y)$. Our randomly initialized network can be viewed as a stack of multiple max-pooling layers, which explains its good performance shown in Figure~\ref{fig:topk-enron random} and Figure~\ref{fig:random-conv}. However, similar analysis is difficult for RNN as an input character passes through the network in many time steps and the influence on edit distance is hard to capture.



\begin{figure*}[!t]	
	\centering
	\includegraphics[width=0.195\textwidth]{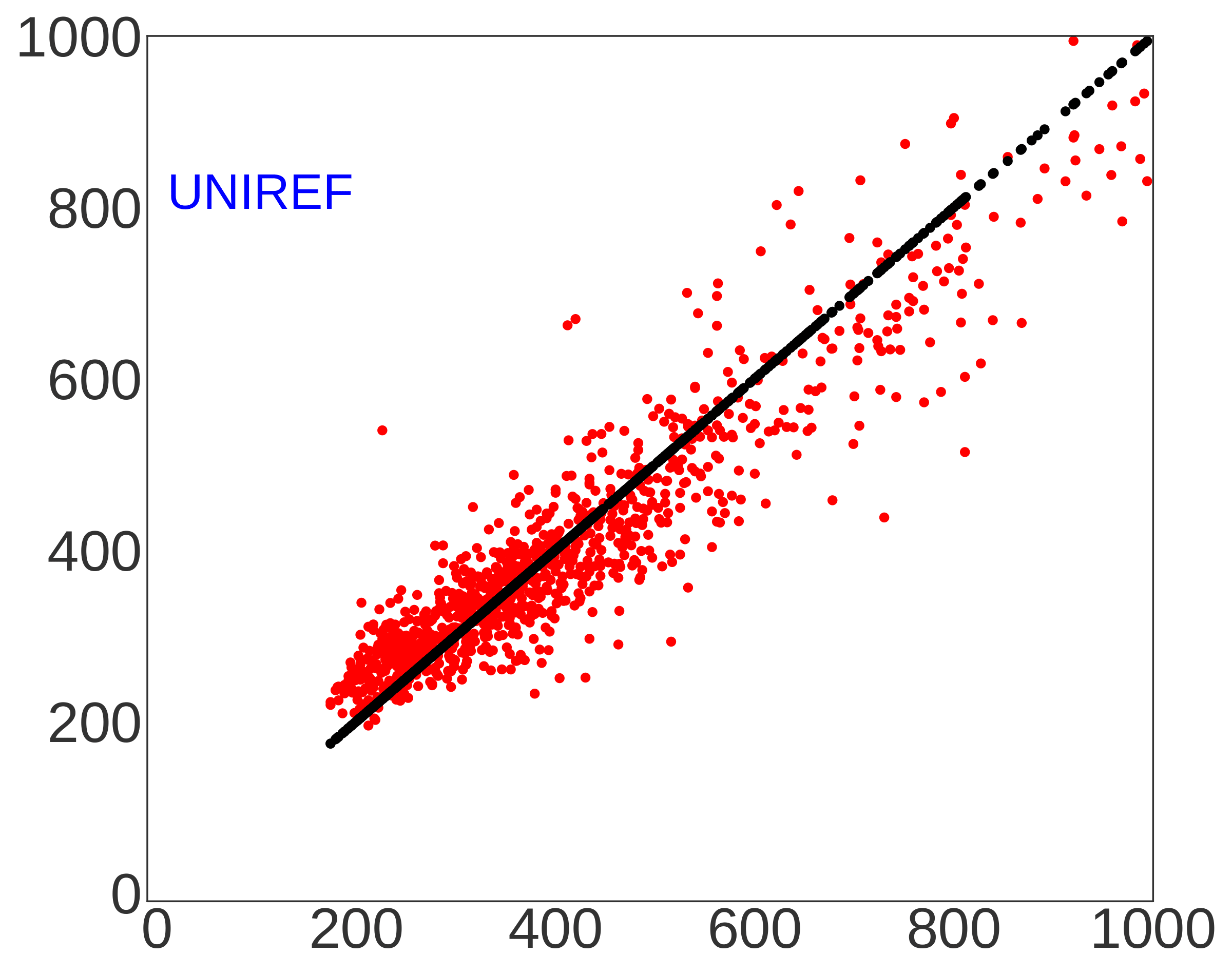}
	\includegraphics[width=0.195\textwidth]{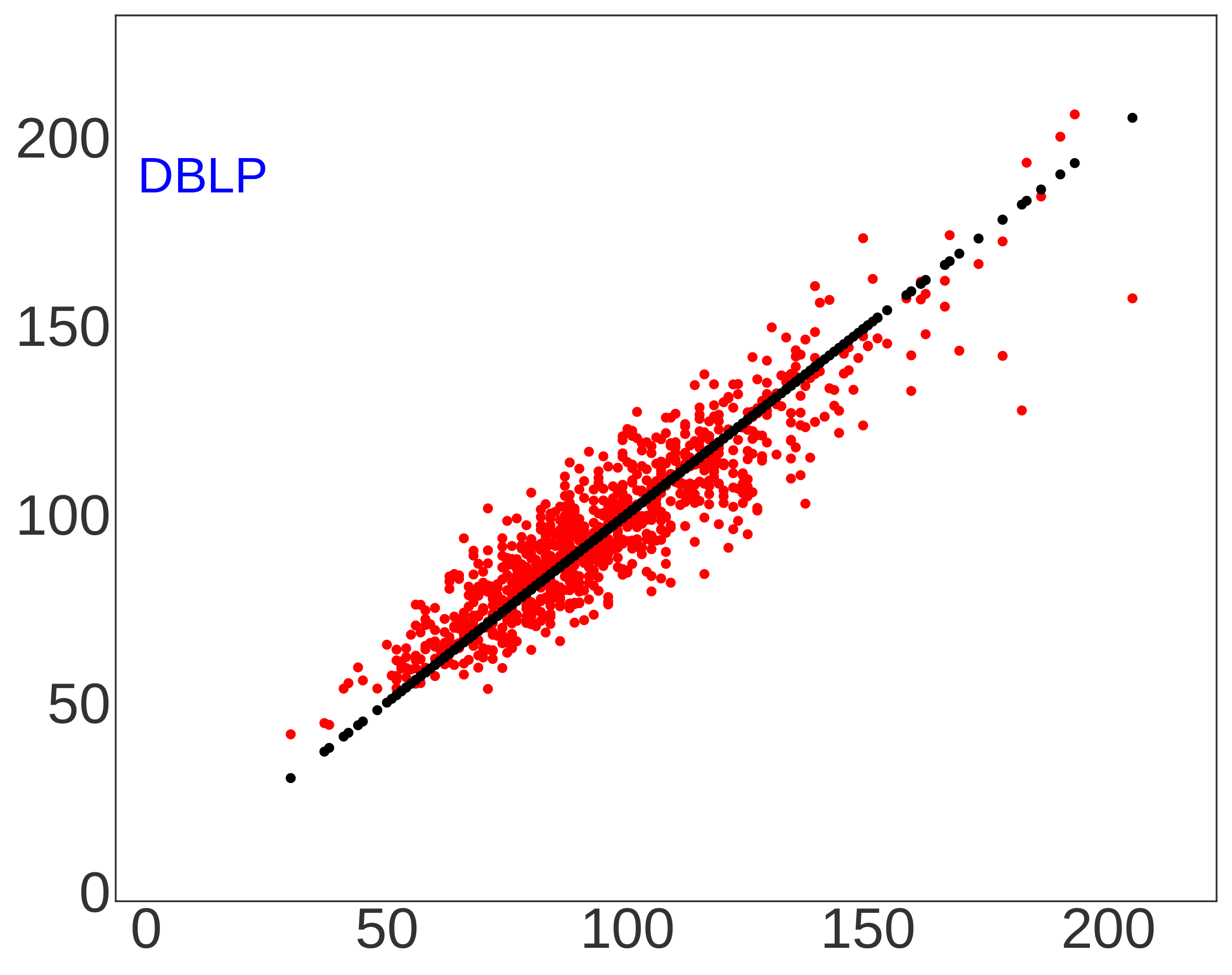}
	\includegraphics[width=0.195\textwidth]{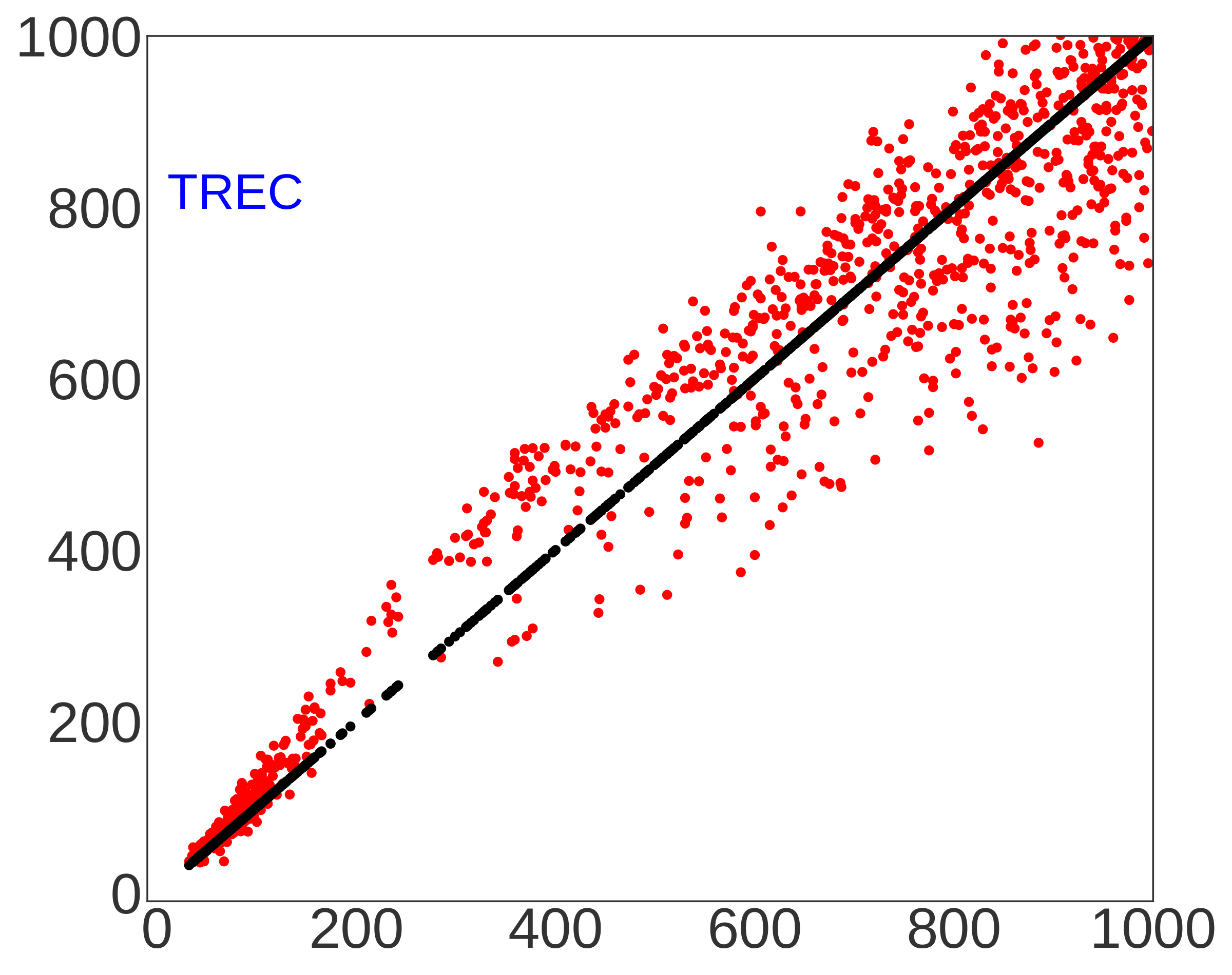}
	\includegraphics[width=0.195\textwidth]{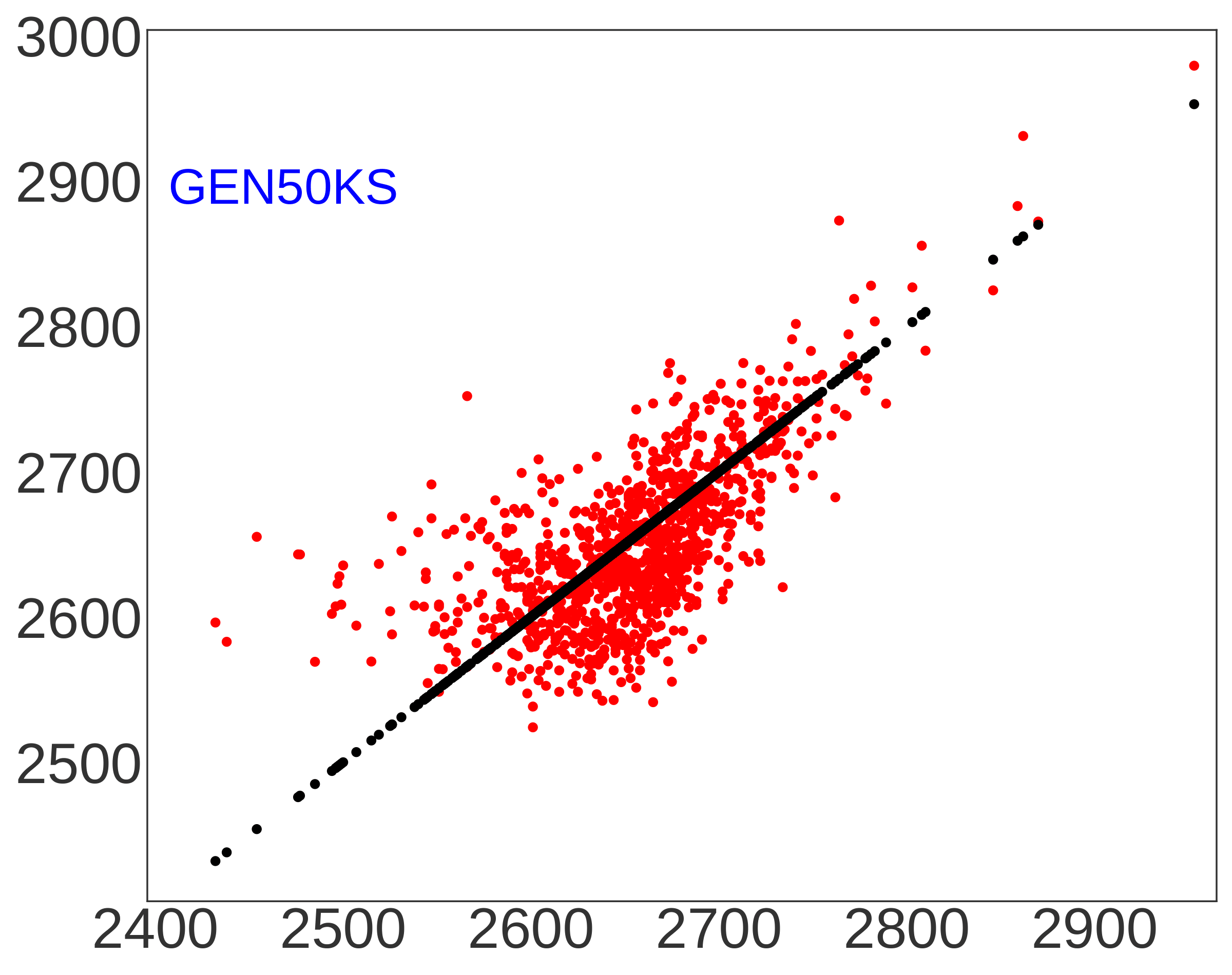}
	\includegraphics[width=0.195\textwidth]{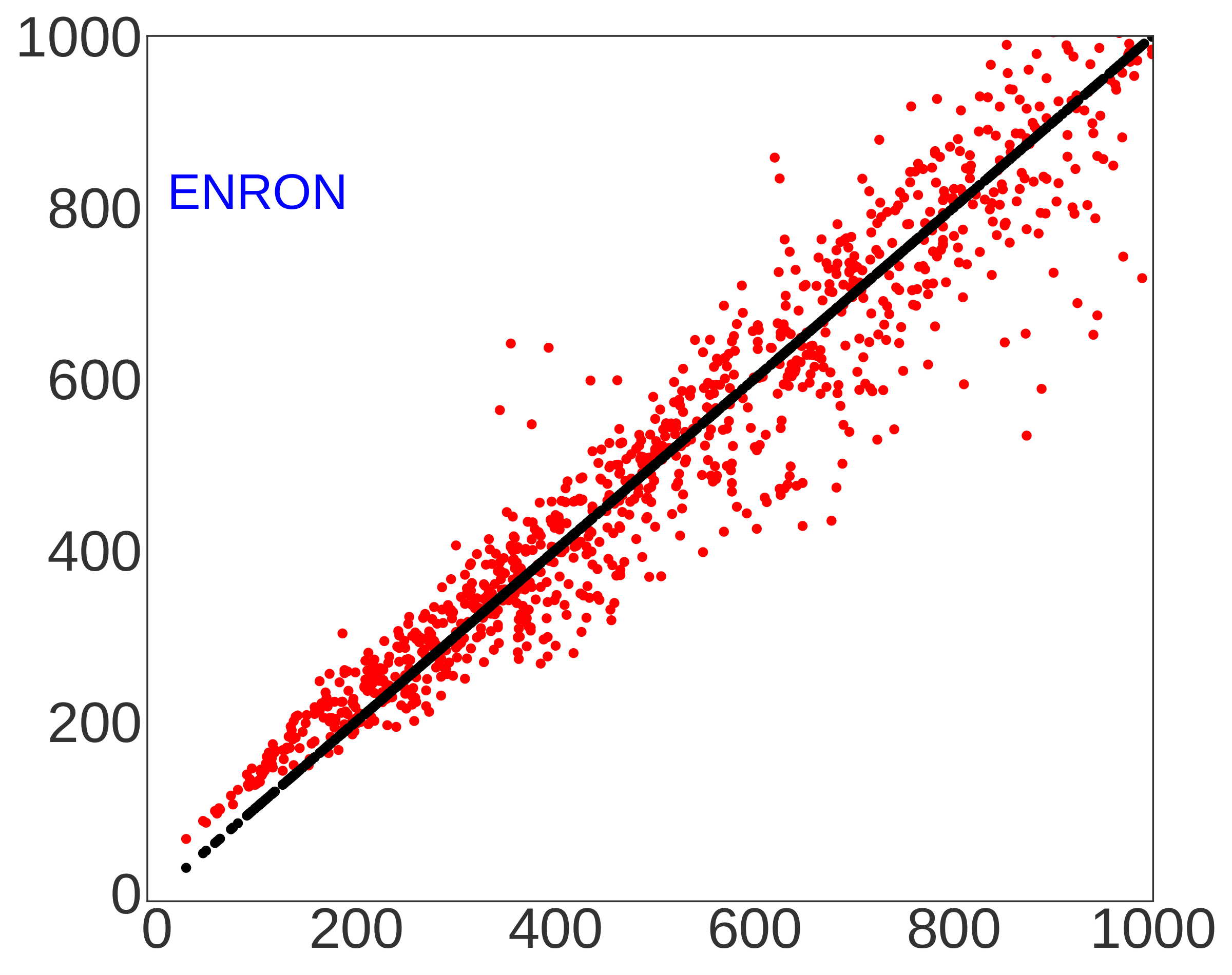}
	\caption{True edit distance (horizontal axis) vs. estimated edit distance (vertical axis) for CNN-ED} 
	\label{fig:distance-estimation}
\end{figure*}

\section{Experimental Results}\label{sec:experiment}

We conduct extensive experiments to evaluate the performance of CNN-ED. Two existing edit distance embedding methods, CGK and GRU, are used as the main baselines. We first introduce the experiment settings, and evaluate the quality of the embeddings generated by CNN-ED. Then, we assess the efficiency of the embedding methods in terms of both computation and storage costs. To demonstrate the benefits of vector embedding, we also test the performance of CNN-ED when used for similarity join and threshold search. Finally, we test the influence of the hyper-parameters (e.g., output dimension, network structure, loss function) on performance. For conciseness, we use CNN to denote CNN-ED in this section. 
The source code is on GitHub\footnote{\url{https://github.com/xinyandai/string-embed}}.  

\begin{table}[]
	\caption{Dataset statistics}
	\label{tab:statistic}
	\begin{center}
		\begin{sf}
			\fontsize{8}{9}\selectfont
			\begin{tabular}{ccccccc}
				\toprule
				DataSet & UniRef & DBLP & Trec & Gen50ks & Enron  \\
				\midrule
				\# Items & 400,000 & 1,385,451 & 347,949 & 50,001 & 245,567 \\
				\midrule
				Avg. Length & 446 & 106 & 845 & 5,000 & 885\\
				\midrule
				Max. Length &  35,214 & 1,627 & 3,948 & 5,153 & 59,420\\
				\midrule
				Alphabet Size &  24 & 37 & 37 & 4 & 37\\
				\bottomrule
			\end{tabular}
		\end{sf}
	\end{center}
\end{table}

\subsection{Experiment Settings}

We conduct the experiments with the fives datasets in Table~\ref{tab:statistic}, which have diverse cardinalities and string lengths. As GRU cannot handle very long strings, we truncated the strings longer than 5,000 in UniRef and Enron to a length of 5,000 following the GRU paper~\cite{gru}. Moreover, as the memory consumption of GRU is too high for datasets with large cardinality, we sample 50,000 item from each dataset for comparisons that involve GRU. In experiments that do not involve GRU, the entire dataset is used. By default, CNN-ED uses 10 one-dimensional convolutional layers with a kernel size of 3 and one linear layer. The dimension of the output embedding is 128.  

\begin{figure*}[!t]	
	\centering
	\includegraphics[width=0.245\textwidth]{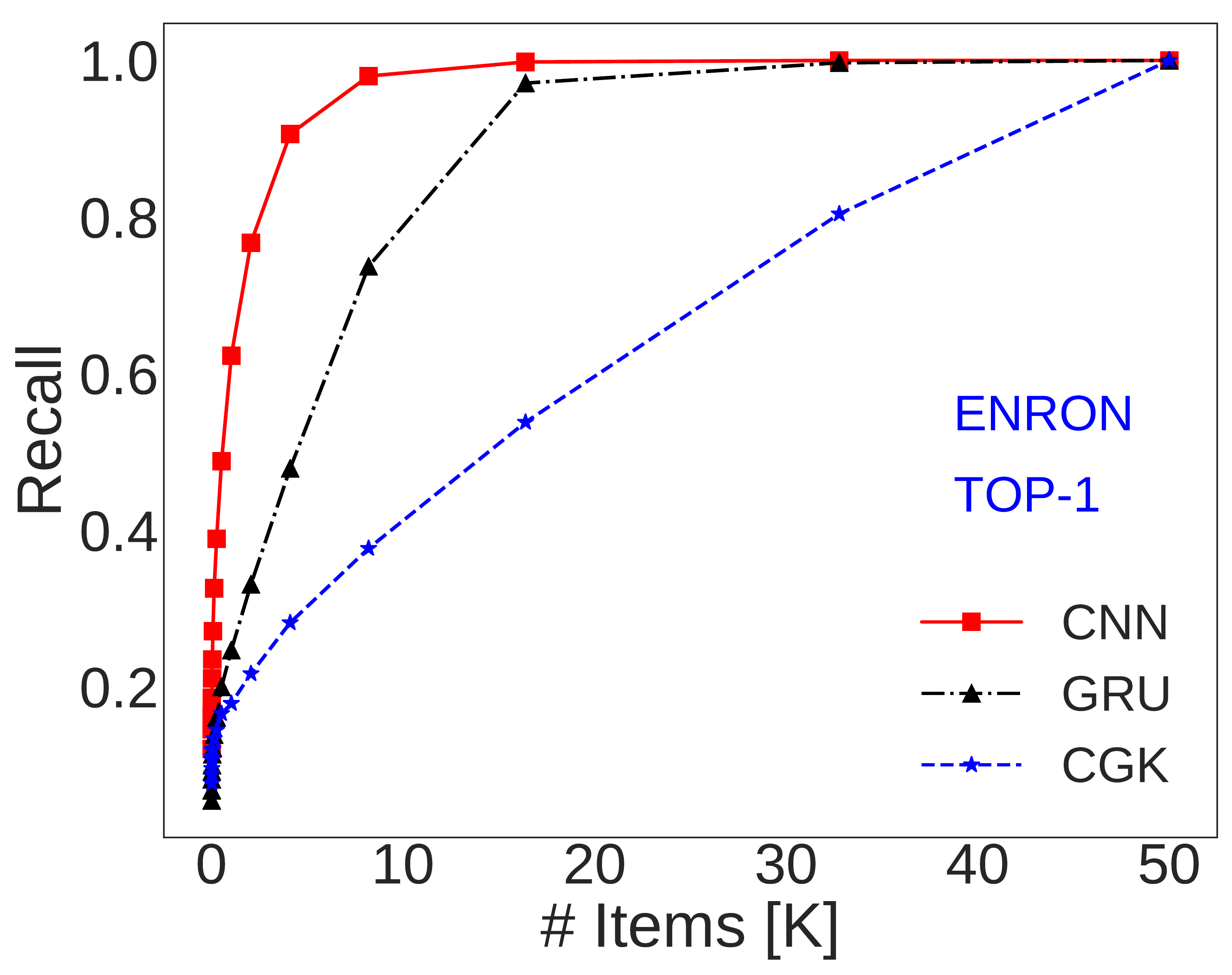}
	\includegraphics[width=0.245\textwidth]{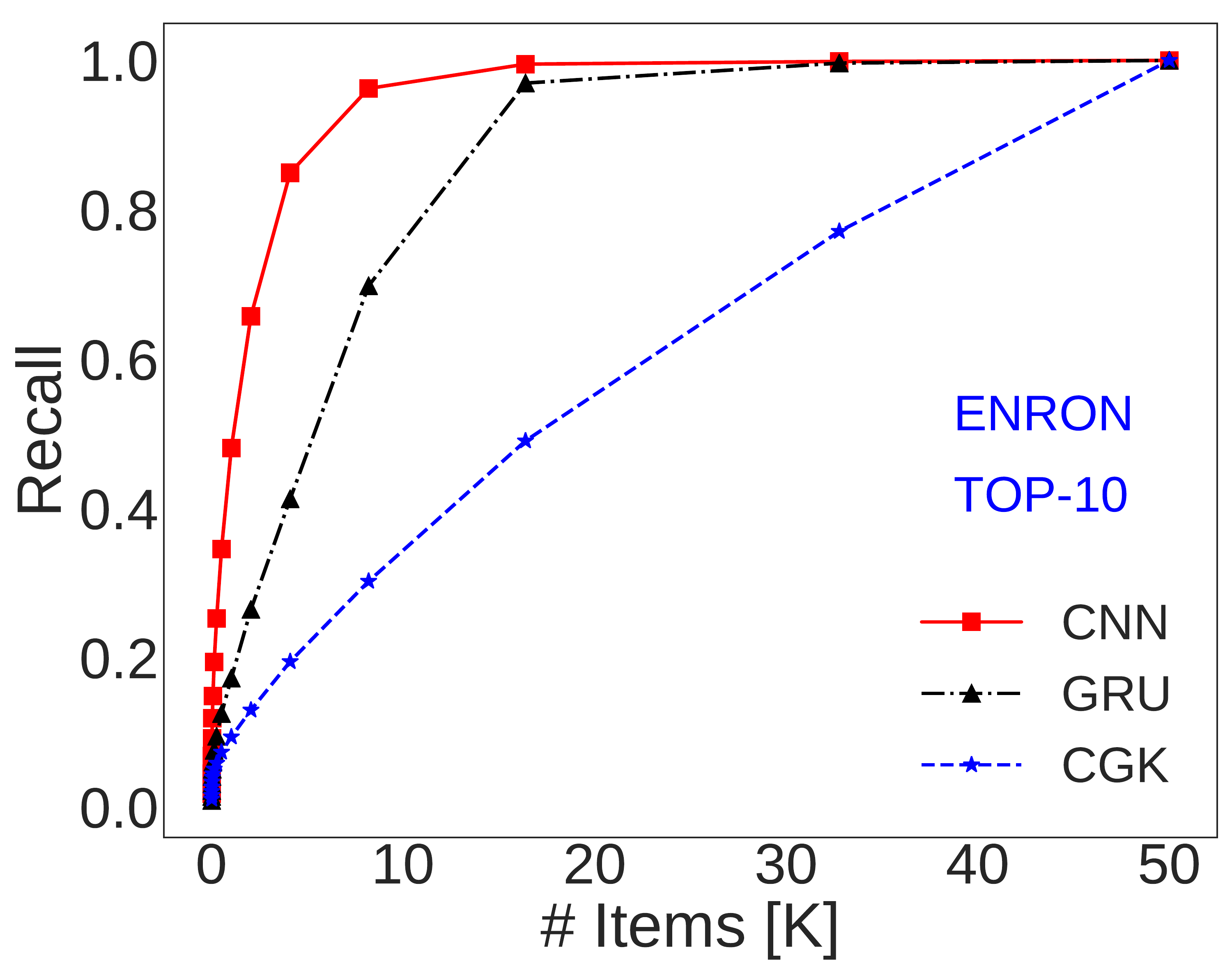} 
	\includegraphics[width=0.245\textwidth]{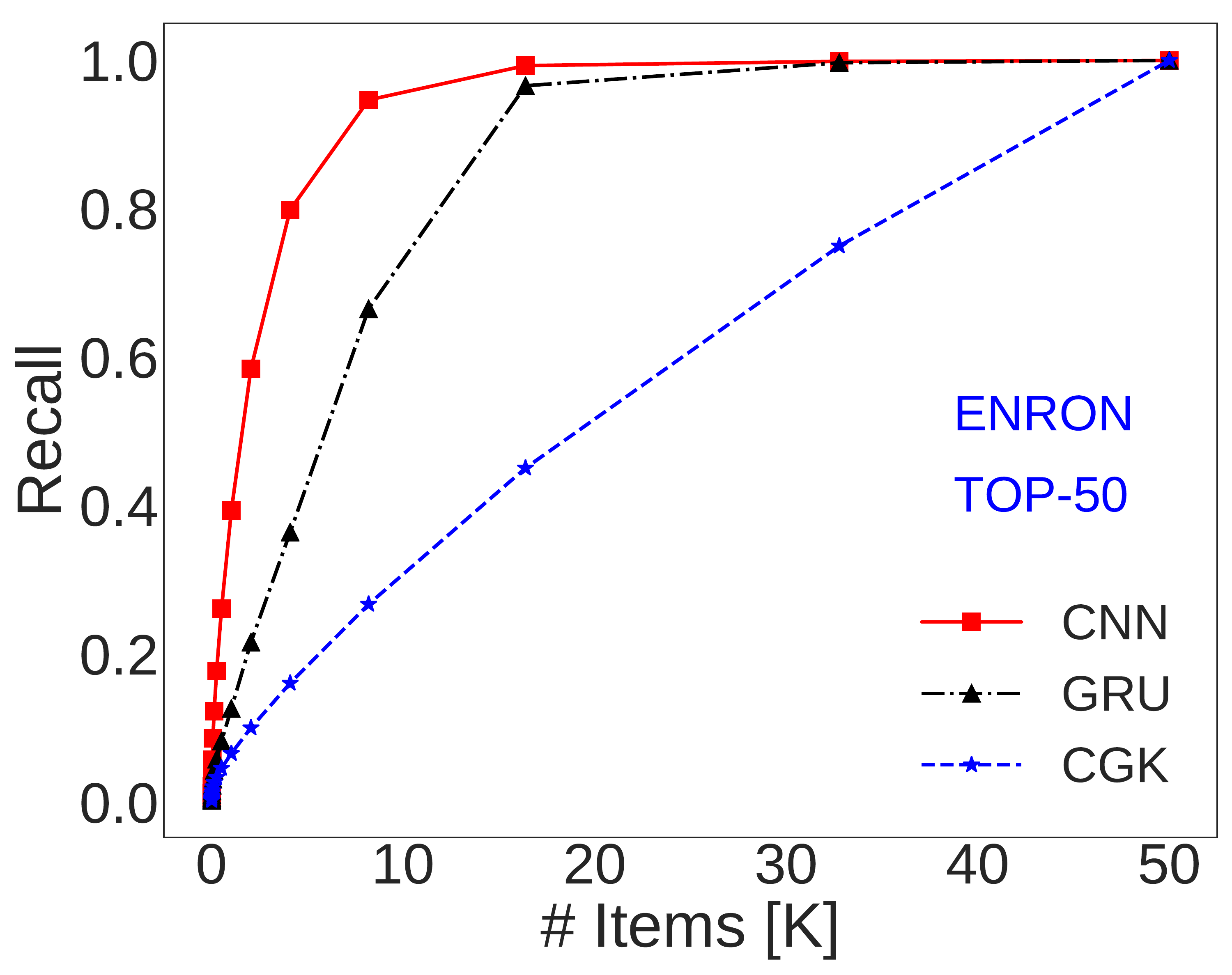} 
	\includegraphics[width=0.245\textwidth]{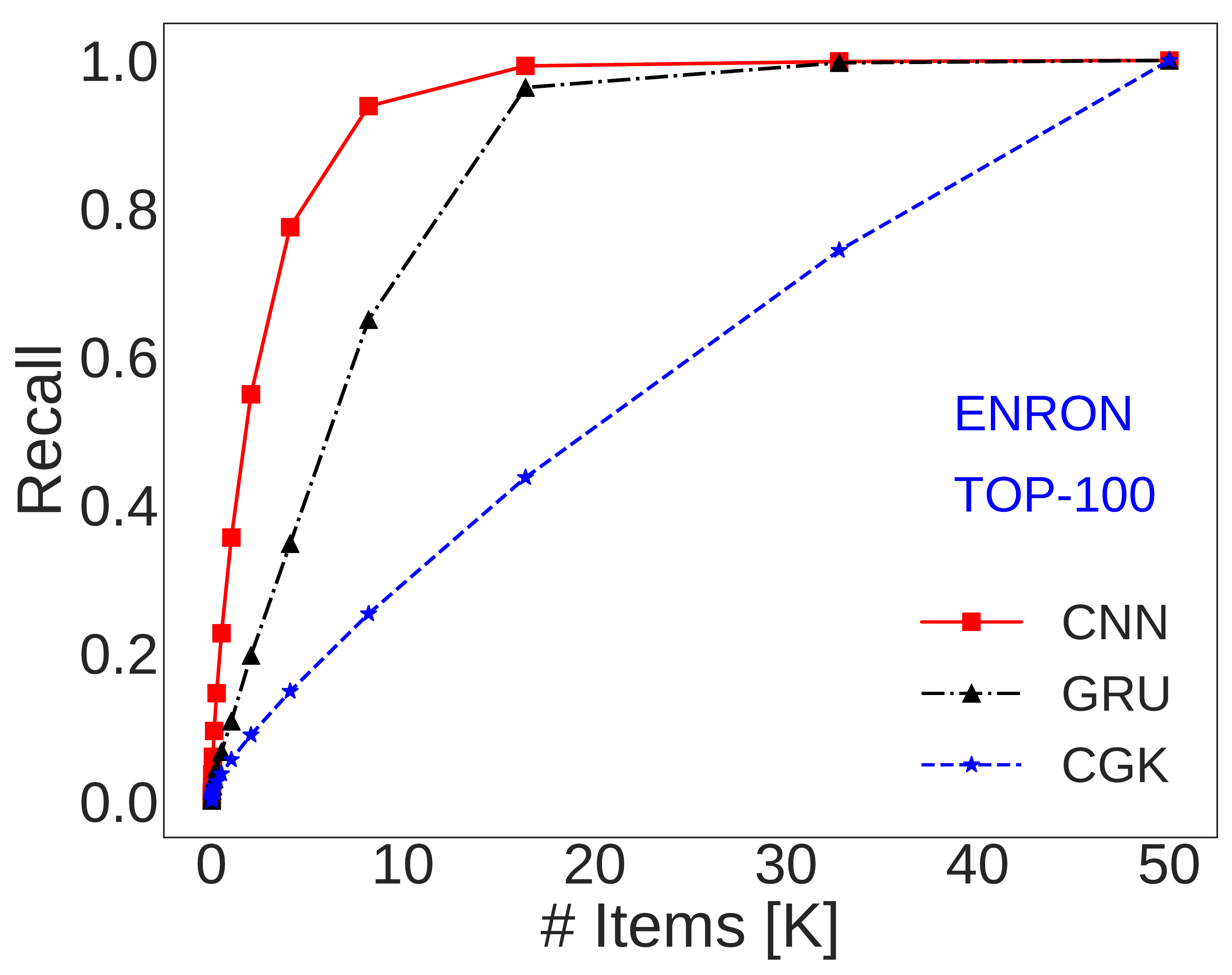} 
	\caption{Recall-item curve comparison among CGK, GRU, CNN for top-$k$ search on the Enron dataset} 
	\label{fig:topk-enron}
\end{figure*}

\begin{figure*}[!t]	
	\centering
	\includegraphics[width=0.245\textwidth]{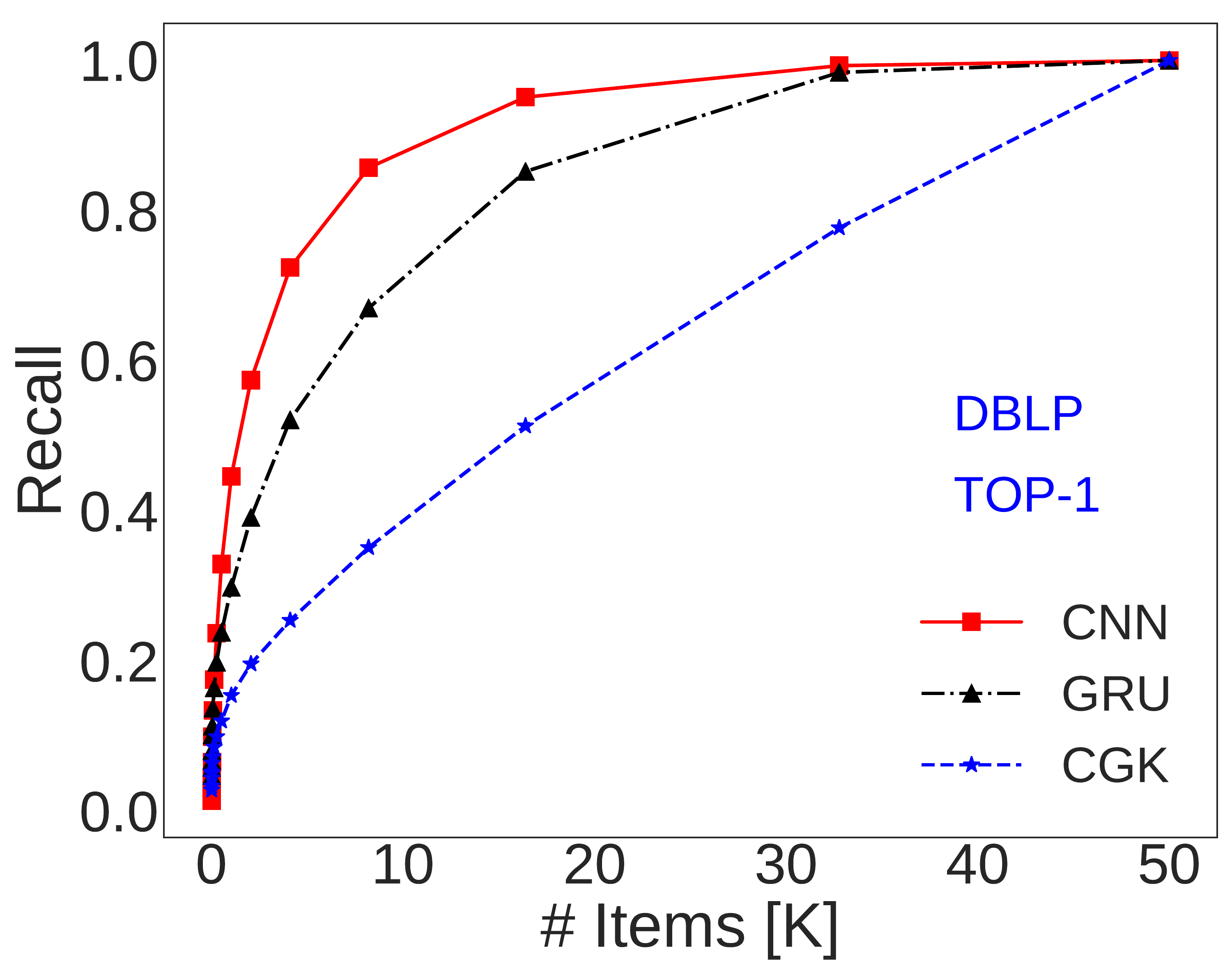}
	\includegraphics[width=0.245\textwidth]{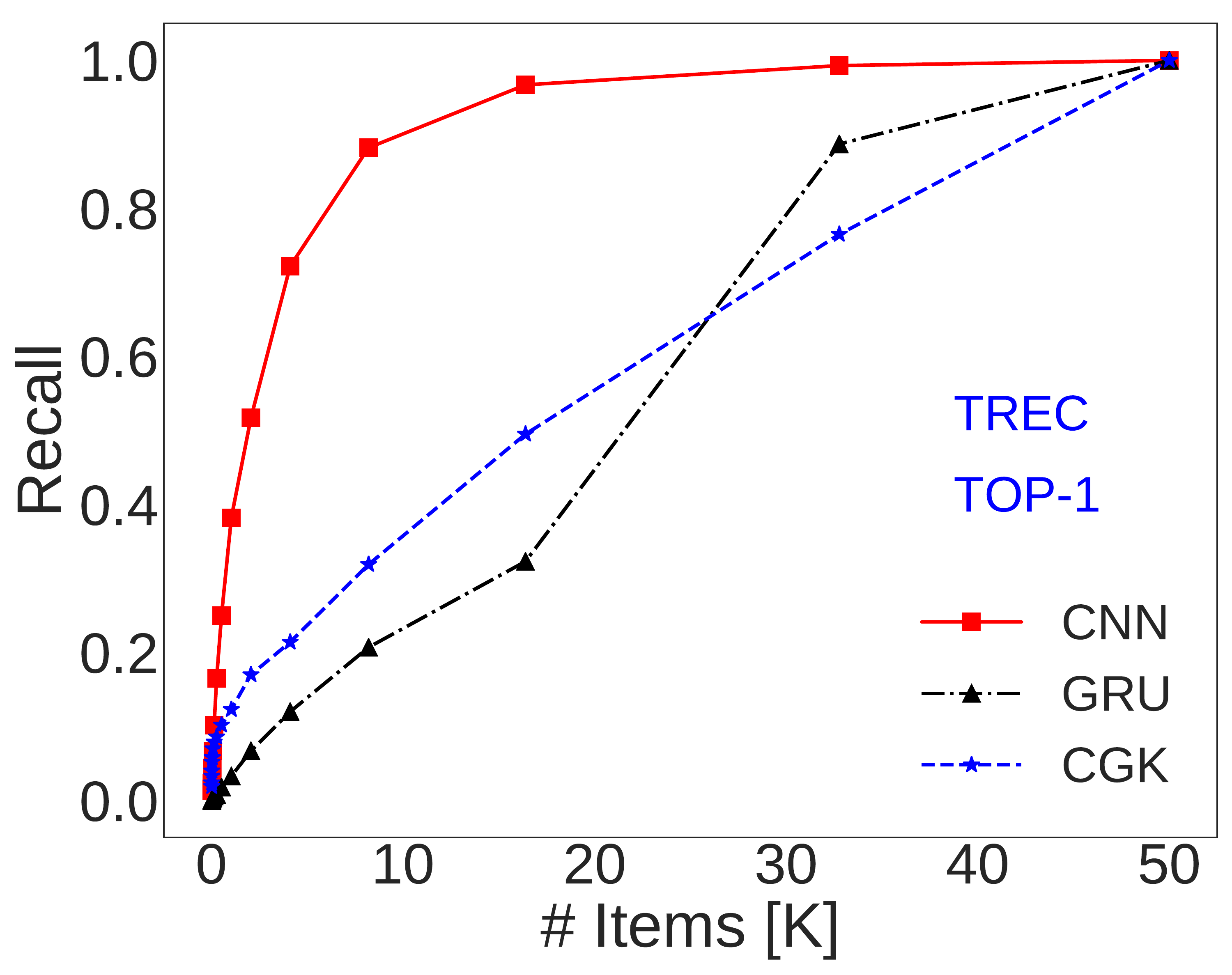}
	\includegraphics[width=0.245\textwidth]{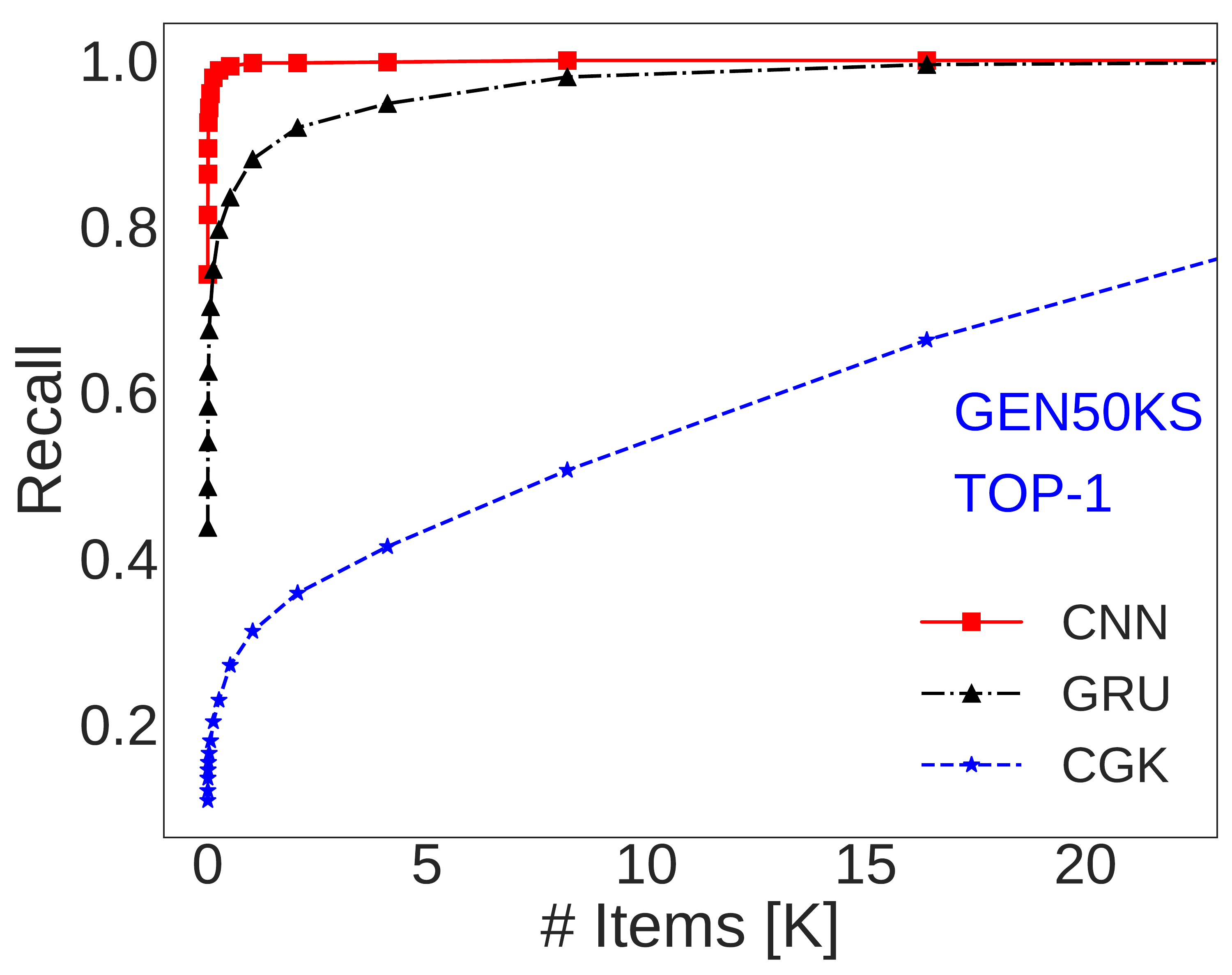}
	\includegraphics[width=0.245\textwidth]{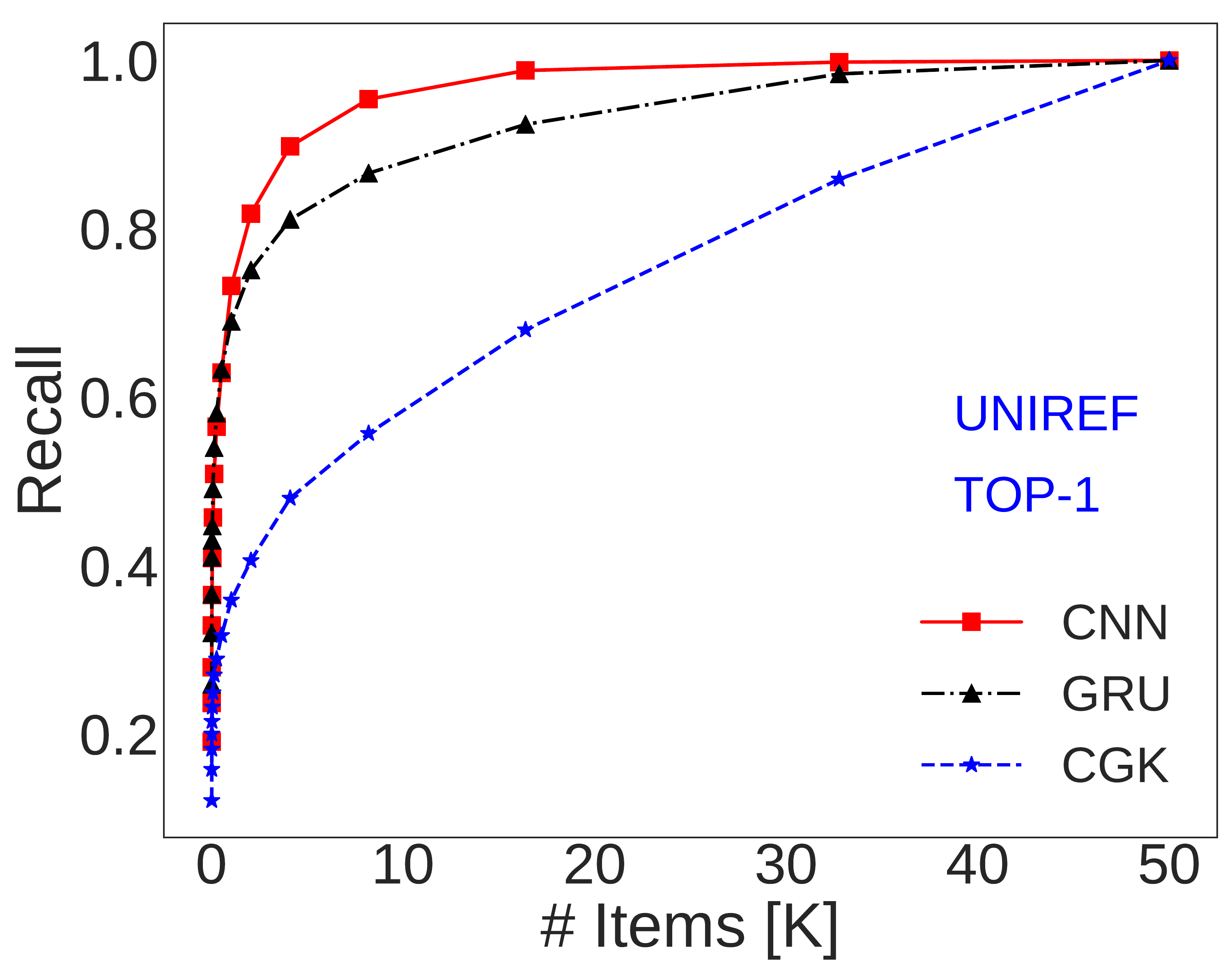}
	\includegraphics[width=0.245\textwidth]{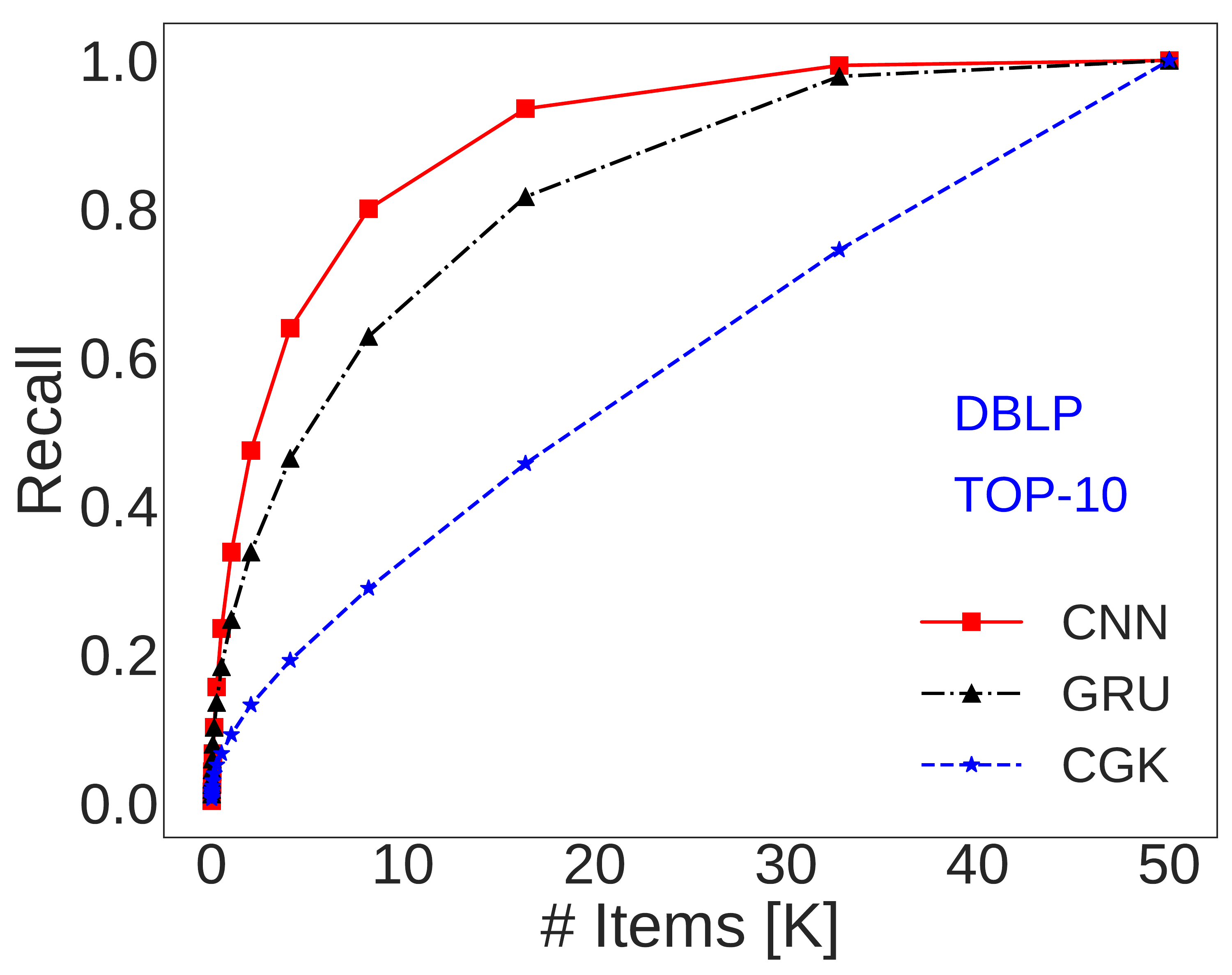}  
	\includegraphics[width=0.245\textwidth]{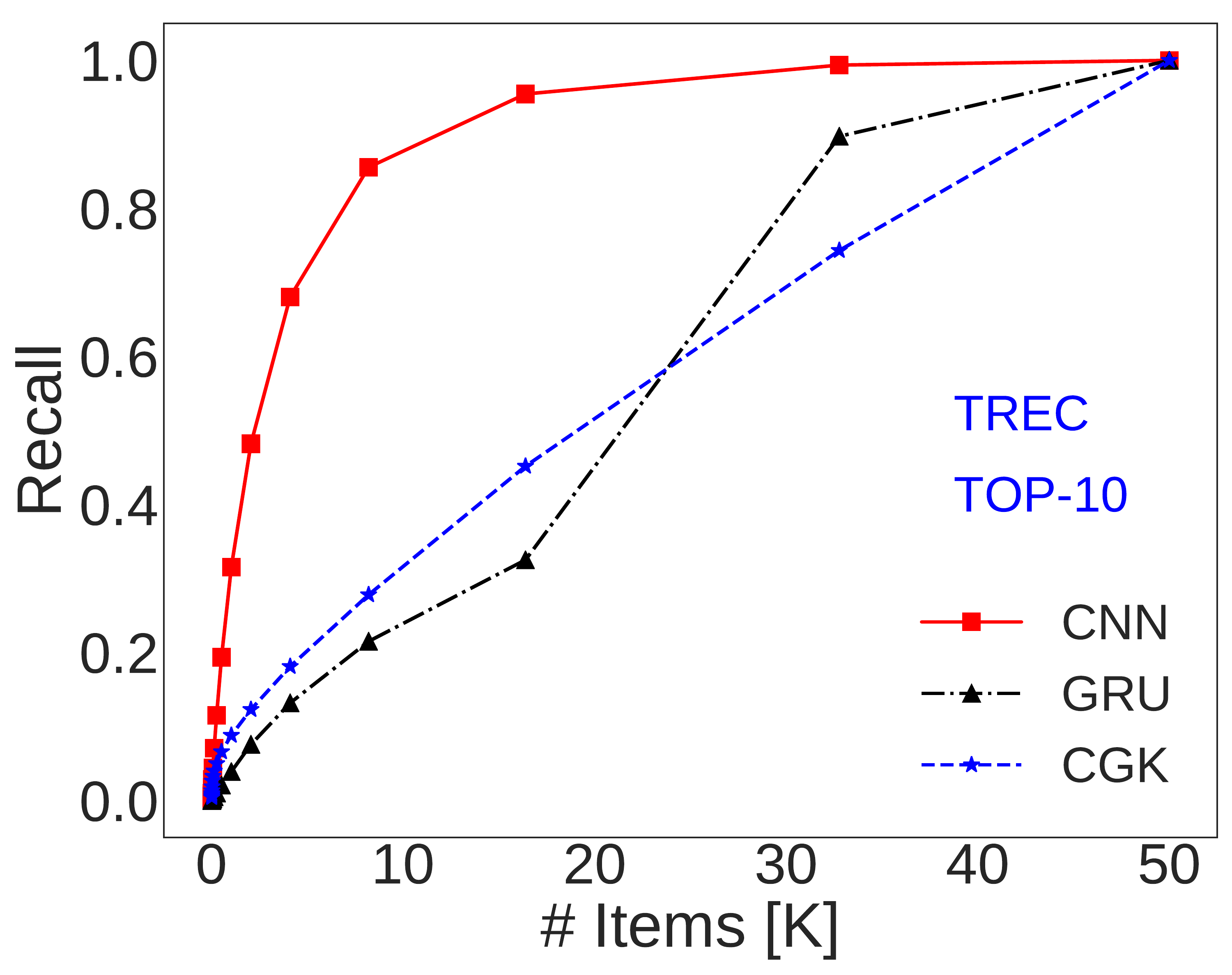} 
	\includegraphics[width=0.245\textwidth]{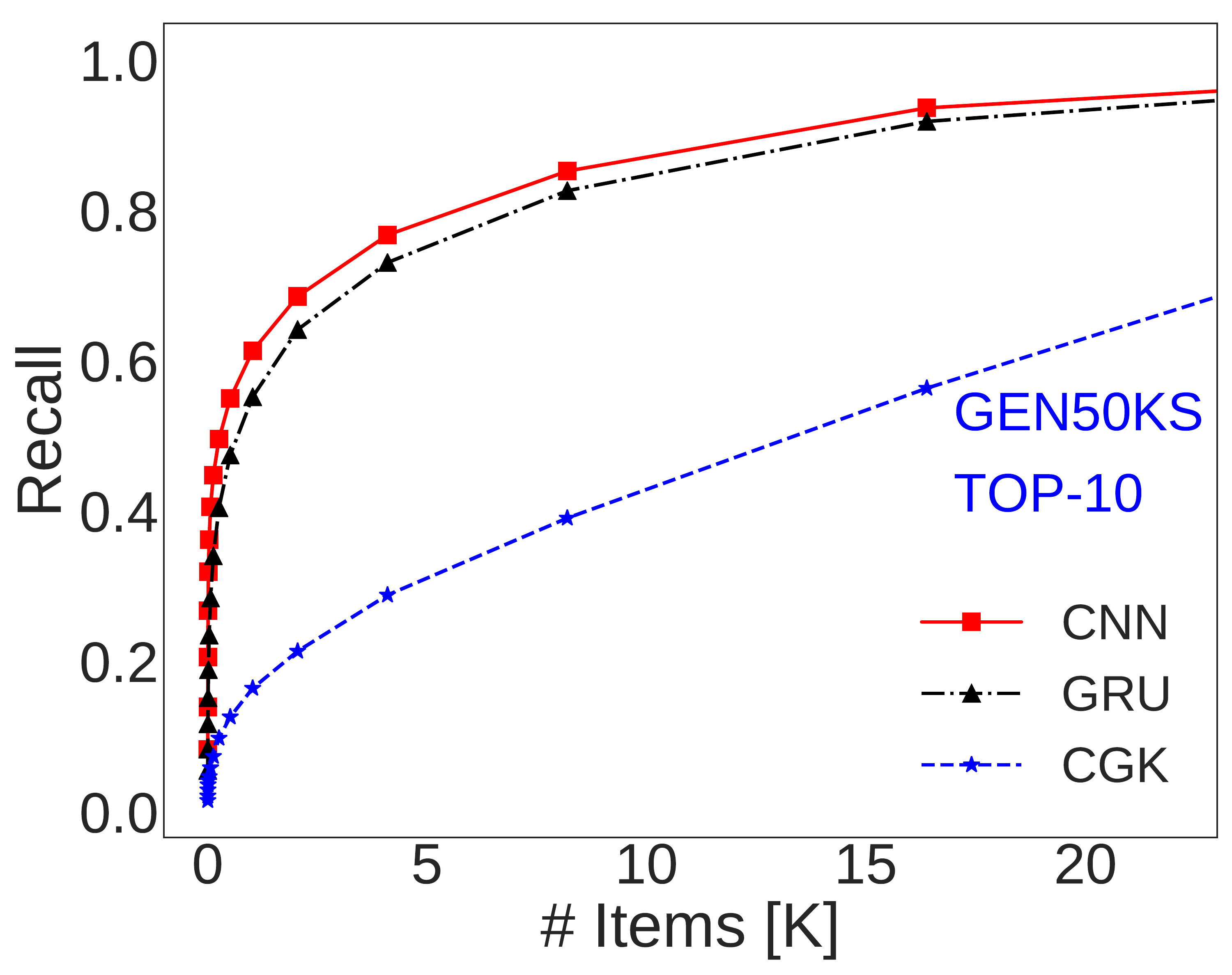}
	\includegraphics[width=0.245\textwidth]{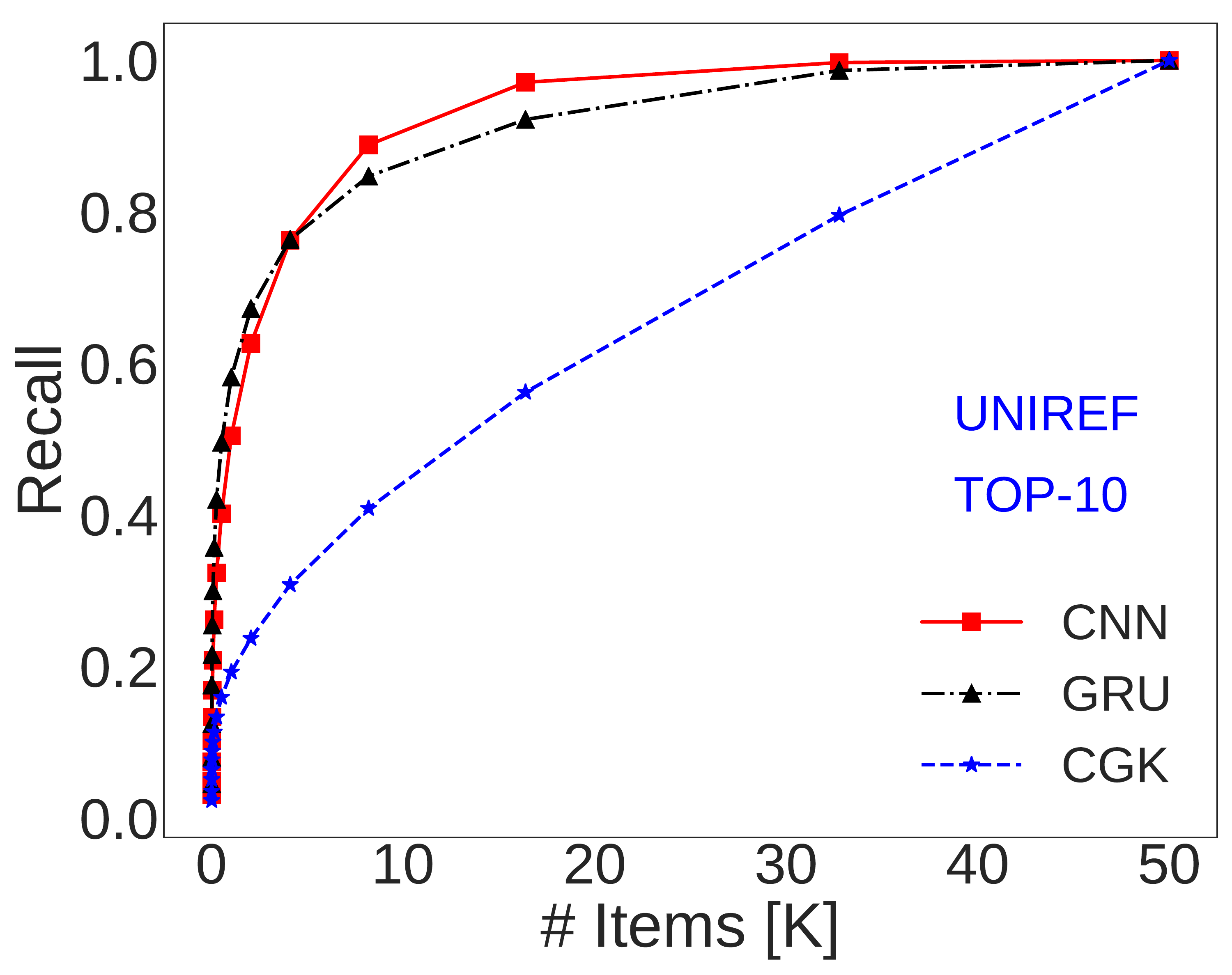}
	\caption{Recall-item curve comparison among CGK, GRU, CNN for top-$k$ search on more datasets} 
	\label{fig:topk-more-dataset}
\end{figure*}     


All experiments are conducted on a machine equipped with GeForce RTX 2080 Ti GPU, 2.10GHz E5-2620 Intel(R) Xeon(R) CPU (16 physical cores), and 48GB RAM. The neural network training and inference experiments are conducted on the GPU while the rest of the experiments are conducted on the CPU. By default, the CPU experiments are conducted using a single thread. For GRU and CNN-ED, we partition each dataset into three disjoint sets, i.e., training set, query set and base set. Both the training set and the query set contain 1,000 items and the other items go to the base set.~\textit{We used only the training set to tune the models and the performance of the models are evaluated on the other two sets.} GRU is trained for 500 epochs as suggested in its code, while CNN-ED is trained for 50 epochs.

\subsection{Embedding Quality}\label{subsec:Embedding Quality}

We assess the quality of the embedding generated by CNN from two aspects, i.e., \textit{approximation error and the ability to preserve edit distance order}.

\begin{table}[]
	\caption{Average edit distance estimation error}
	\label{tab:distance-estimation}
	\begin{center}
		\begin{sf}
			\fontsize{8}{9}\selectfont
			\begin{tabular}{ccccccc}
				\toprule
				DataSet & UniRef & DBLP & Trec & Gen50ks & Enron  \\
				\midrule
				CGK  &0.590  & 63.602  &6.856   & 0.452&  0.873  \\
				\midrule
				GRU  &0.275  & 0.175 &  46.840  & 0.419 &  0.126  \\
				\midrule
				CNN  &  \textbf{0.125} & \textbf{0.087}  &\textbf{0.141}  & \textbf{ 0.401}&  \textbf{0.123}\\
				\bottomrule
			\end{tabular}
		\end{sf}
	\end{center}
\end{table}

To provide an intuitive illustration of the approximation error of the CNN embeddings, we plot the true edit distance and the estimated edit distance of 1,000 randomly sampled query-item pairs in Figure~\ref{fig:distance-estimation}. The estimated edit distance of a string pair $(s_i, s_j)$ is computed using a linear function of the Euclidean distance $\Vert f(s_i)-f(s_j)\Vert$. The linear function is introduced to account for possible translation and scaling between the two distances, and it is fitted on the training set without information from the base and query set. The results show that the distance pairs locate closely around the $y=x$ line (the black one), which suggests that CNN embeddings provide good edit distance approximation.

To quantitatively compare the approximation error of the embedding methods, we report the average edit distance estimation error in Table~\ref{tab:distance-estimation}. The estimation error for a string pair is defined as $e=\frac{|g(d(f(s_i), f(s_j))-\ED(s_i,s_j)|}{\ED(s_i,s_j)}$, in which $\ED(s_i,s_j)$ is the true edit distance and $g(d(f(s_i), f(s_j))$ is the  edit distance estimated from embeddings. The distance function $d(f(s_i), f(s_j))$ is Hamming distance for CGK and Euclidean distance for GRU and CNN. $g(\cdot)$ is a function used to calculate edit distance using distance in the embedding space, and it is fitted on the training set. We set $g(\cdot)$ as a linear function for GRU and CNN, and a quadratic function for CGK as the theoretical guarantee of CGK in Equation~\eqref{equ:CGK} has a quadratic form. The reported estimation error is the average of all possible query-item pairs. The results show that CNN has the smallest estimation error on all five datasets, while overall CGK has the largest estimation error. This is because CGK is data-independent, while GRU and CNN use machine learning to fit the data. The performance of GRU is poor on the Trec dataset and similar phenomenon is also reported in its original paper~\cite{gru}.            

To evaluate the ability of the embeddings to preserve edit distance order, we plot the recall-item curve in Figure~\ref{fig:topk-enron} and Figure~\ref{fig:topk-more-dataset}. The recall-item curve is widely used to evaluate the performance of metric embedding. To plot the curve, we first find the top-$k$ most similar strings for each query in the base set using linear scan. Then, for each query, items in the base set are ranked according to their distance to the query in the embedding space. If the items ranking top $T$ contain $k'$ of the true top-$k$ neighbors, the recall is $k'/k$. For each value of $T$, we report the average recall of the 1,000 queries. Intuitively, a good embedding should ensure that a neighbor with a high rank in edit distance (i.e., having smaller edit distance than most items) also has a high rank in embedding distance. In this case, the recall is high for a relatively small $T$. The results show that CNN consistently outperforms CGK and GRU on all five datasets and for different values of $k$. The recall-item performance also agrees with the estimation error in Table~\ref{tab:distance-estimation}. CNN has the biggest advantage in estimation error on Trec and its item-recall performance is also significantly better than CGK and GRU on this dataset. On Gen50ks, GRU and CNN have similar estimation error, and the item-recall performance of CNN is only slightly better than GRU.               

\subsection{Embedding Efficiency}

We compare the efficiency of the embedding algorithms from various aspects in Table~\ref{table:efficiency}. \textit{Train time} is the time to train the model on the training set, and \textit{embed time} is the average time to compute the embedding for a string (also called inference). \textit{Compute time} is the average time to compute the distance between a pair of strings in the embedding space. \textit{Embed size} is the memory consumption for storing the embeddings of a dataset, and \textit{Raw} is the size of the original dataset. Note that the embed time of GRU and CNN is measured on GPU, while the embed time of CGK is measured on CPU as the CGK algorithm is difficult to parallelize on GPU. For GRU, the embedding size of the entire dataset is estimated using a sample of 50,000 strings.

The results show that CNN is more efficient than GRU in all aspects. CNN trains and computes string embedding at least 2.6x and 13.7x faster than GRU, respectively. Moreover, CNN takes more than 290x less memory to store the embedding, and computes distance in the embedding space over 400x faster. When compared with CGK, CNN also has very attractive efficiency. CNN computes approximate edit distance at least 14x faster than CGK and uses at least an order of magnitude less memory. We found that CNN is more efficient than GRU and CGK mainly because it has much smaller output dimension. For example, on the Gen50ks dataset, the output dimensions of CGK and GRU are 322x and 121x of CNN, respectively. Note that even with much smaller output dimension, CNN embedding still provides more accurate approximation for edit distance than CGK and GRU, as we have shown in Section~\ref{subsec:Embedding Quality}. CGK embeds strings faster than both GRU and CNN as the two learning-based models need to conduct neural network inference while CGK follows a simple random procedure.

\begin{table}[ht]
	\caption{Embedding efficiency comparison}
	\label{table:efficiency}
	\begin{center}
		\fontsize{7.2}{9}\selectfont
		\begin{tabular}{c|ccccccc}
			\toprule
			DataSet& Method & UniRef & DBLP & Trec & Gen50ks & Enron  \\
			\hline
			\multicolumn{1}{c|}
			{\multirow{2}{*}{ \makecell{Train \\ Time (s)} }}
			&GRU & 31.8 & 13.2 & 26.3 & 31.3 & 34.9  \\
			\multicolumn{1}{c|}{}
			&\textbf{CNN} & \textbf{4.31} &  \textbf{4.96} &  \textbf{5.19} &  \textbf{1.61} &  \textbf{5.63} \\
			
			\hline
			\multicolumn{1}{c|}
			{\multirow{3}{*}{\makecell{Embed \\ Time \\($\mu$s)}}}
			&CGK   & \textbf{52.2} & \textbf{16.2} & \textbf{56.6} & \textbf{105.6} & \textbf{63.6} \\
			
			\multicolumn{1}{c|}{}
			&GRU  & 8332 & 2340 & 7654 & 12067 & 7650 \\
			\multicolumn{1}{c|}{}
			&\textbf{CNN}  &  378.8 & 134.8 & 361.8 & 172.2 & 548.4 \\
			
			\hline
			\multicolumn{1}{c|}
			{\multirow{3}{*}{\makecell{Comp. \\ Time \\ ($\mu$s)}}}
			&CGK   &1.72 & 0.60 & 1.36 & 1.65 & 1.71 \\
			
			\multicolumn{1}{c|}{}
			&GRU  &  123.7 & 47.2 & 129.1 & 18.0 & 177.7 \\
			
			\multicolumn{1}{c|}{}
			&\textbf{CNN($10^{-2}$)}  &\textbf{4.6} & \textbf{4.2} & \textbf{4.2} & \textbf{4.2} & \textbf{4.5}\\
			
			\hline
			\multicolumn{1}{c|}
			{\multirow{4}{*}{\makecell{Embed \\ Size}}}
			&Raw  & {170MB} & {140MB} & 280MB & 238MB & 207MB \\
			\multicolumn{1}{c|}{}
			&CGK  & {5.59GB} & {6.45GB} & 4.86GB & 0.70GB & 3.43GB \\
			\multicolumn{1}{c|}{}
			&GRU  & 372GB & 621GB & 378GB & 7GB & 338GB \\
			\multicolumn{1}{c|}{}
			&\textbf{CNN}  &\textbf{195MB} &\textbf{676MB} & \textbf{169MB} & \textbf{24MB} & \textbf{119MB}  \\
			\bottomrule
		\end{tabular}
	\end{center}
\end{table}

\subsection{Similarity Search Performance}

In this part, we test the performance of CNN when used for the two string similarity search problems discussed in Section~\ref{sec:background}, \textit{threshold search and similarity join}.  

For threshold search, model training and dataset embedding are conducted before query processing. When a query comes, we first calculate its embedding, and then use the distances in the embedding space to rank the items, and  finally conduct exact edit distance computation in the ranked order. Following~\cite{embedjoin}, the thresholds for UniRef, DBLP, Trec, Gen50ks and Enron are set as 100, 40, 40, 100 and 40, respectively. We compare with HSsearch, which supports threshold search with a hierarchical segment index. For CNN, we measure the average query processing time when reaching certain recall, where recall is defined as the number of returned similar string pairs over the total number of ground truth similar pairs. 

\begin{table}[]
	\caption{Average query time for threshold search (in ms)}
	\label{tab:threshold-search}
	\begin{center}
		\begin{sf}
			\fontsize{8}{9}\selectfont
			\begin{tabular}{ccccccc}
				\toprule
				DataSet & UniRef & DBLP & Trec & Gen50ks & Enron  \\
				\midrule
				HSsearch & 4333  &6907  &222  &393  & 76 \\
				\midrule
				CNN(R=0.6) &26  &263  &37  &1.73 &12 \\
				\midrule
				CNN(R=0.8) &66  &478  & 44 &1.74  &13 \\
				\midrule
				CNN(R=0.9) &143   &1574  &58  &1.74  &15 \\
				\midrule
				CNN(R=0.95) &254   &2296  &80  &1.75  &15 \\
				\midrule
				CNN(R=0.99) &1068   &3560  &93  &1.77  &21 \\
				\midrule
				CNN(R=1) &  3007   &4321   &116   & 1.79 & 22\\
				\bottomrule
			\end{tabular}
		\end{sf}
	\end{center}
\end{table}

The results in Table~\ref{tab:threshold-search} show that when approximate search is acceptable, CNN can achieve very significant speedup over HSsearch. At a recall of 0.6, the speedup over HSsearch is at least 6x and can be as much as 227x. In principle, CNN is not designed for exact threshold search as there are errors in its edit distance approximation. However, it also outperforms HSsearch when the recall is 1, which means all ground truth similar pairs are returned, and the speedup is at least 1.44x for the five datasets. 

\begin{table}[h]
	\caption{Average query time for GRU and CNN (in ms)}
	\label{tab:threshold-search-gru}
	\begin{center}
		\begin{sf}
			\fontsize{8}{9}\selectfont
			\begin{tabular}{ccccccc}
				\toprule
				Recall & 0.6 & 0.8 & 0.9 & 0.95 & 0.99 & 1  \\
				\midrule
				CNN & 5.3  & 6.7  & 8.6  &14.3  & 60.2 & 91.0 \\
				\midrule
				GRU &  2980.5   &  3012.1   & 3059.6  & 3059.6 & 3590.2 &3590.2\\
				\bottomrule
			\end{tabular}
		\end{sf}
	\end{center}
\end{table}

To demonstrate the advantage of the accurate and efficient embedding provided by CNN, we compare CNN and GRU for threshold search in Table~\ref{tab:threshold-search-gru}. The dataset is a sample with 50,000 items from the DBLP dataset (different from the entire DBLP dataset used in Table~\ref{tab:threshold-search}). We conduct sampling because the GRU embedding for the whole dataset does not fit into memory. The results show that CNN can be up to 500x faster than GRU for attaining the same recall. Detailed profiling finds that distance computation is inefficient with long GRU embedding (as shown in Table~\ref{table:efficiency}) and CNN embedding better preserve the order of edit distance (as shown in Figure~\ref{fig:topk-more-dataset}).

\begin{figure}[!t]	
	\centering
	\begin{subfigure}{0.48\linewidth}
		\centering 
		\includegraphics[width=\textwidth]{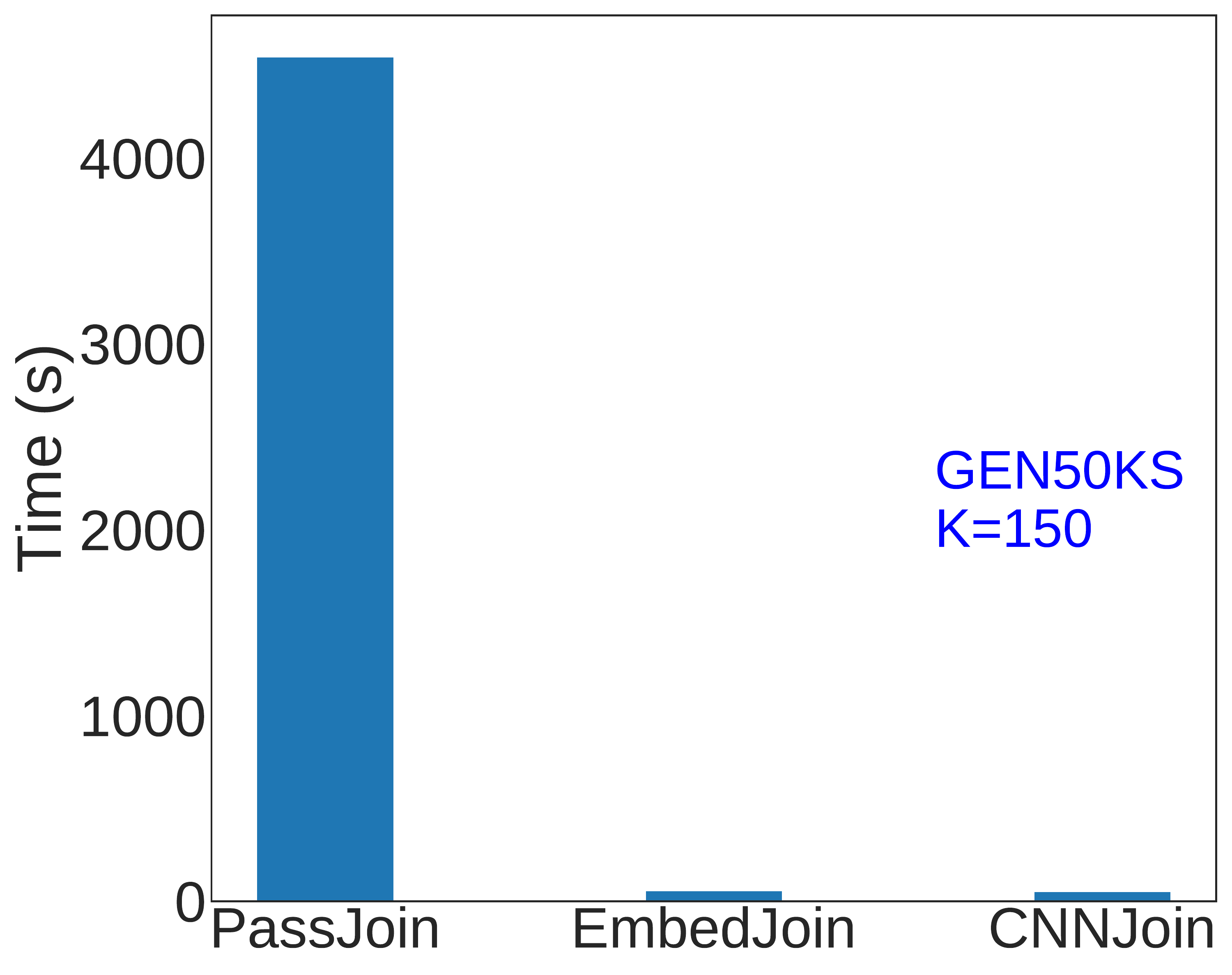} 
		\caption{Gen50ks}
		\label{fig:GEN}
	\end{subfigure}
	\begin{subfigure}{0.48\linewidth}
	\centering 
		\includegraphics[width=\textwidth]{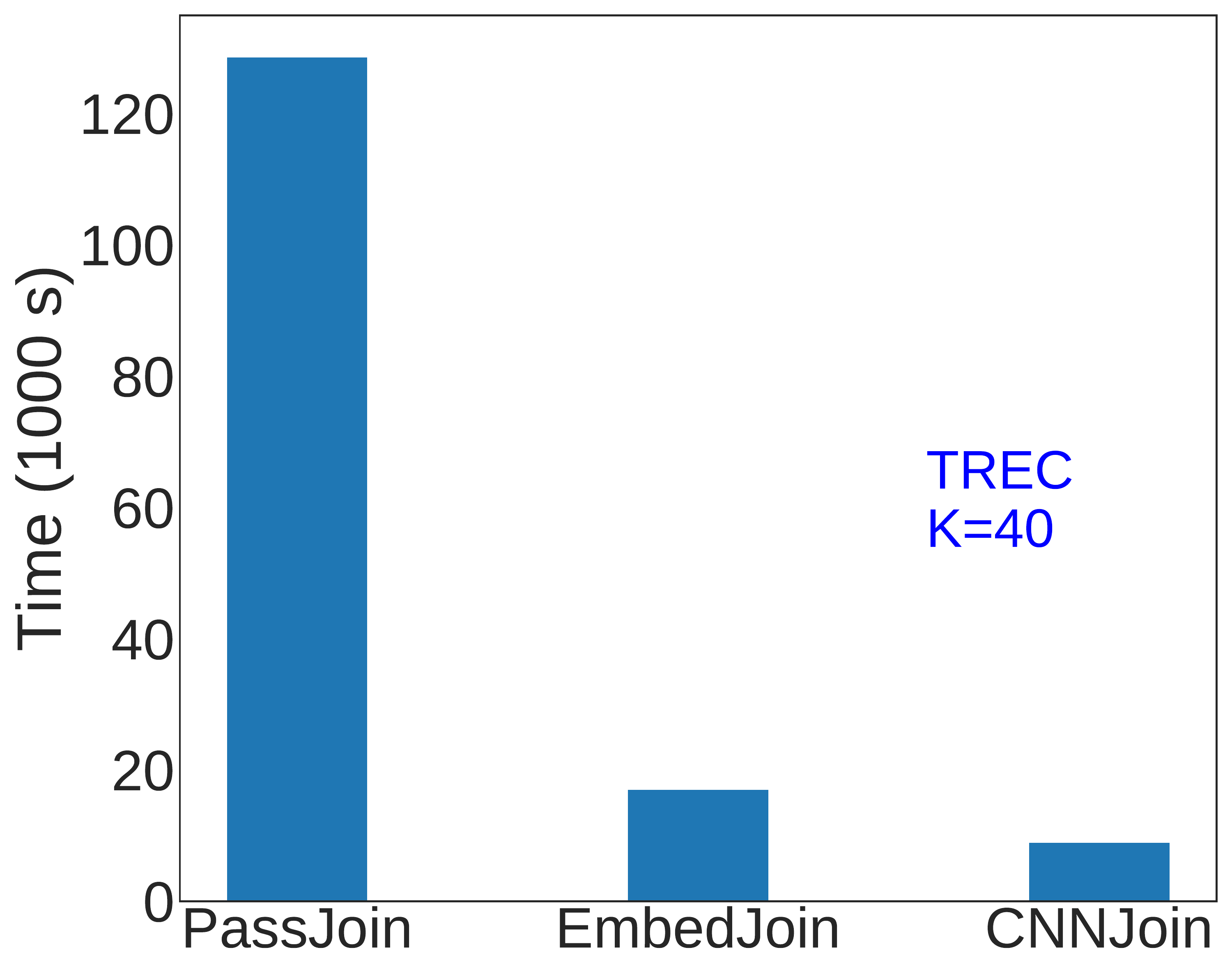} 
		\caption{Trec}
		\label{fig:xx}
	\end{subfigure}
	\caption{Time comparison for similarity join} 
	\label{fig:join-time}
\end{figure}

We compare CNN with EmbedJoin and PassJoin for similarity join in Figure~\ref{fig:join-time}. PassJoin partitions a string into a set of segments and creates inverted index for the segments, then generates similar string pairs using the inverted index. EmbedJoin uses the CGK embedding and locality sensitive hashing (LSH) to filter unnecessary edit distance computations, which is the state-of-the-art method for string similarity join. Note that PassJoin is an exact method while EmbedJoin is an approximate method. Different from threshold search, the common practice for similarity join is to report the end-to-end time, which includes both pre-processing time (e.g., index building in EmbedJoin) and edit distance computation time. Therefore, we include the time for training the model and embedding the dataset in the results of CNN. For Gen50ks and Trec, the thresholds for similar string pairs are set as 150 and 40, respectively, following the EmbedJoin paper. For EmbedJoin and CNN, we report the time taken to reach a recall of 0.99.

The results show that EmbedJoin outperforms CNN on the Gen50ks dataset but CNN performs better than EmbedJoin on the Trec dataset. To investigate the reason behind the results, we decompose the running time of CNN into training time, embedding time and search time in Table~\ref{tab:join}. On the smaller Gen50ks dataset (with 50,000 items), CNN takes 160.1s, 8.6s and 48.8s for training, embedding and search, respectively, while EmbedJoin takes 52.8 seconds in total. The results suggest that CNN performs poorly on Gen50ks because the dataset is small and the long training time cannot be amortized. On the larger Trec dataset (with 347,949 items), the training time (and embedding time) is negligible (only 5\% of the total time) and CNN is 1.76x faster than EmbedJoin due to its the high quality embedding. Therefore, we believe CNN will have a bigger advantage over EmbedJoin on larger dataset. We have tried to run the algorithms on the DBLP dataset but both PassJoin and EmbedJoin fail.  

This set of experiments shows that CNN embeddings provide promising results for string similarity search. We believe that more sophisticated designs are possible to further improve performance, e.g., using multiple sets of independent embeddings to introduce diversity, combining with Euclidean distance similarity methods such as vector quantization and proximity graph, and incorporating the various pruning rules used in existing string similarity search work.

\begin{table}[]
	\caption{Time decomposition for similarity join (in seconds)}
	\label{tab:join}
	\begin{center}
		\begin{sf}
			\fontsize{7}{8}\selectfont
			\begin{tabular}{cccccc}
				\toprule
				Dataset	 & CNN-Train & CNN-Embed &  CNN-Search & EmbedJoin   \\
				\midrule
				Gen50ks &  160.1   &  8.6   & 48.8 & 52.8 \\
				\midrule
				Trec &  510.7  & 190.2 & 8924.6  & 16944.0  \\
				\bottomrule
			\end{tabular}
		\end{sf}
	\end{center}
\end{table}

\begin{figure}[!t]	
	\centering
	\begin{subfigure}{0.48\linewidth}
		\centering 
		\includegraphics[width=\linewidth]{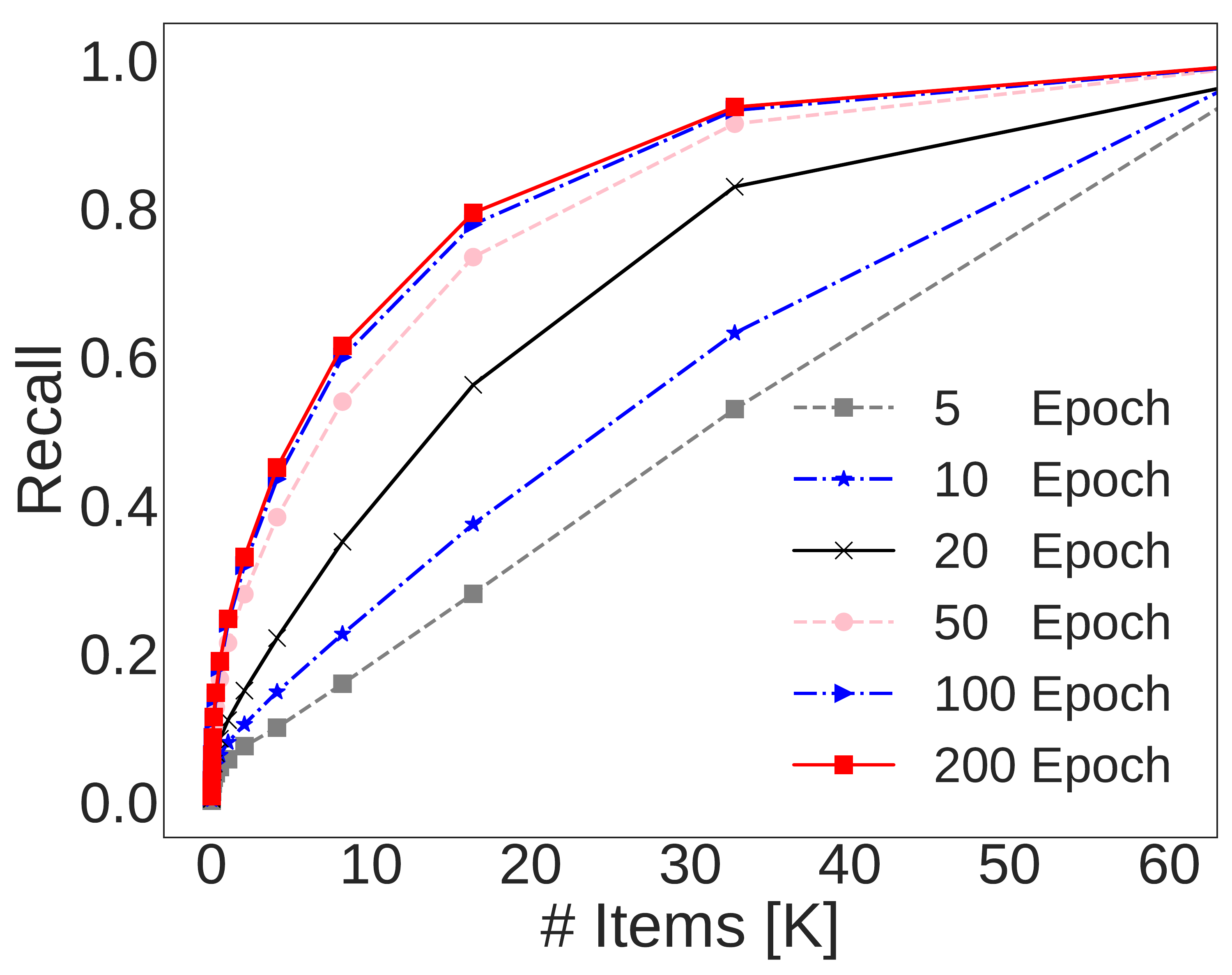} 
		\caption{Epoch count}
		\label{fig:epoch}
	\end{subfigure}
	\begin{subfigure}{0.48\linewidth}
		\centering 
		\includegraphics[width=\linewidth]{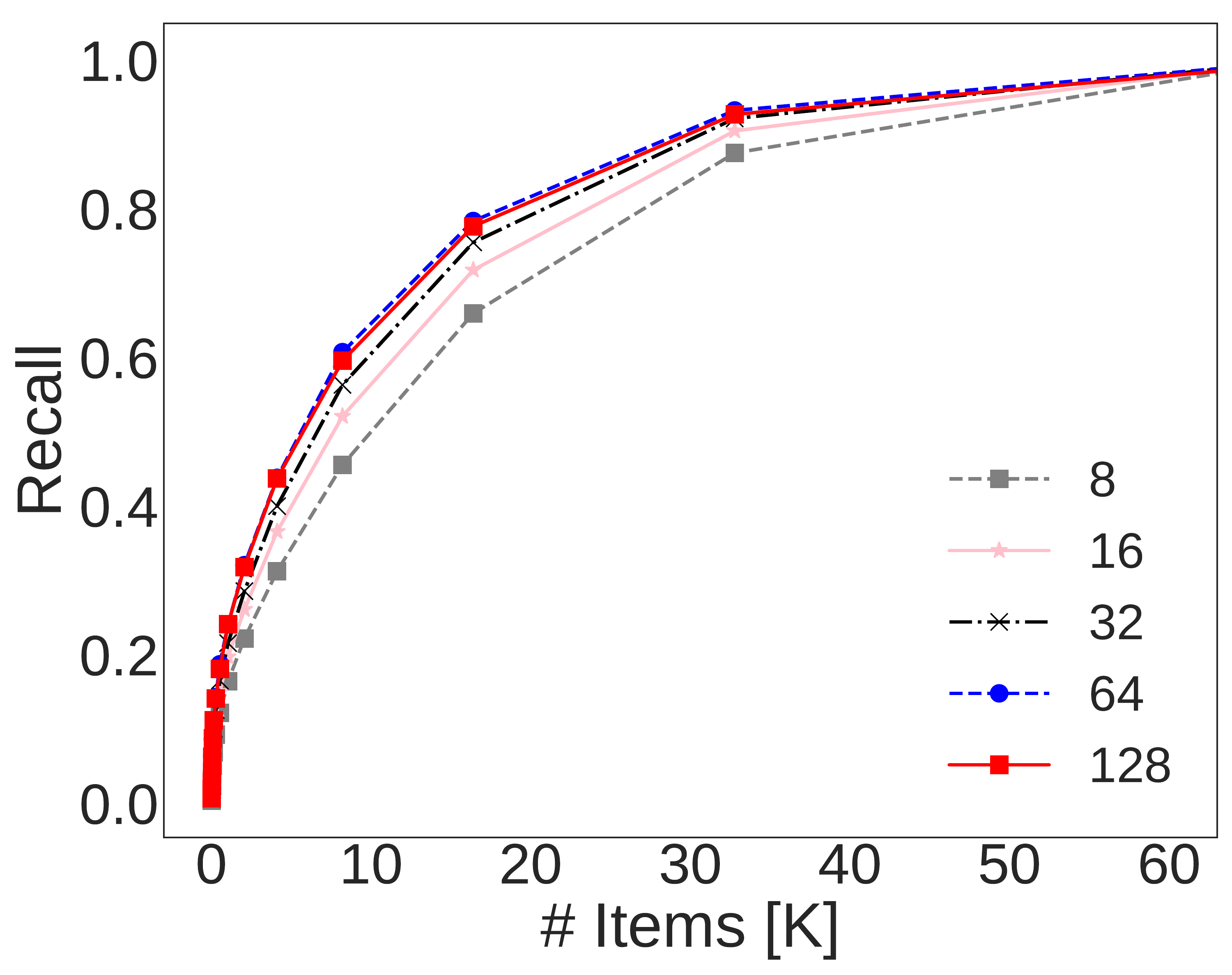} 
		\caption{Output dimension}
		\label{fig:dim}
	\end{subfigure}\\
	\begin{subfigure}{0.48\linewidth}
		\centering 
		\includegraphics[width=\linewidth]{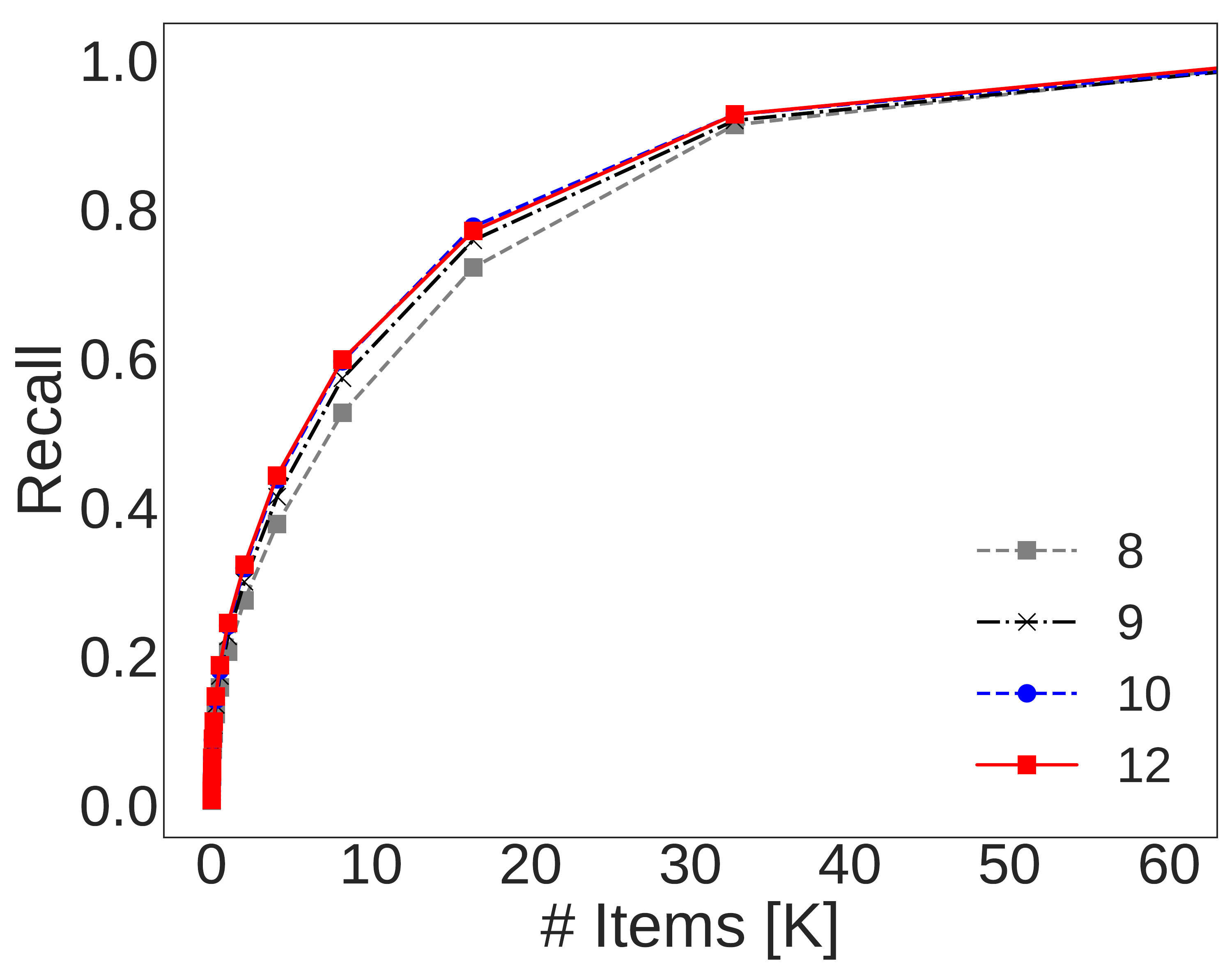} 
		\caption{Number of layer}
		\label{fig:layer}
	\end{subfigure}
	\begin{subfigure}{0.48\linewidth}
		\centering 
		\includegraphics[width=\linewidth]{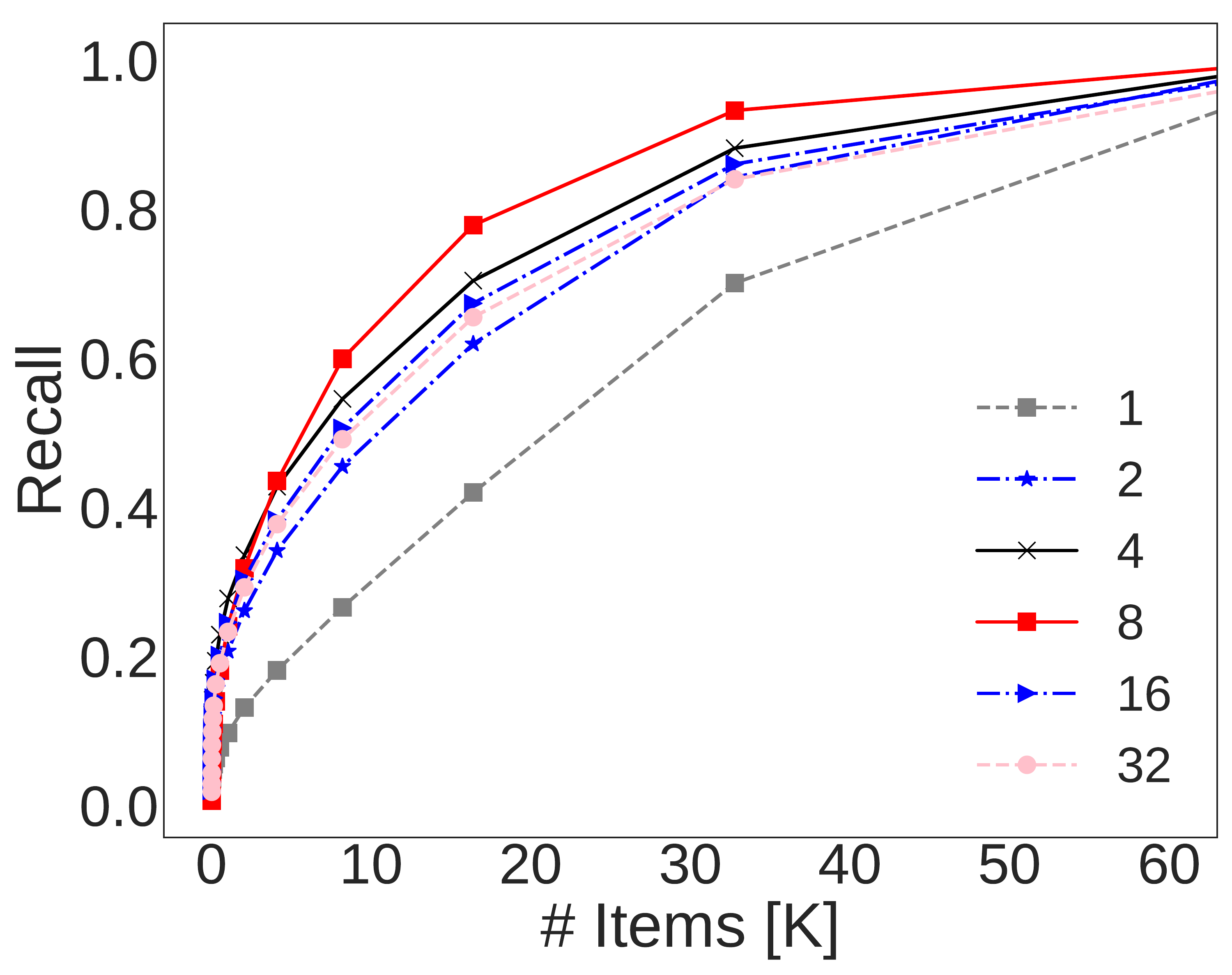} 
		\caption{Number of  kernel}
		\label{fig:channel}
	\end{subfigure}
	\begin{subfigure}{0.48\linewidth}
		\centering 
		\includegraphics[width=\linewidth]{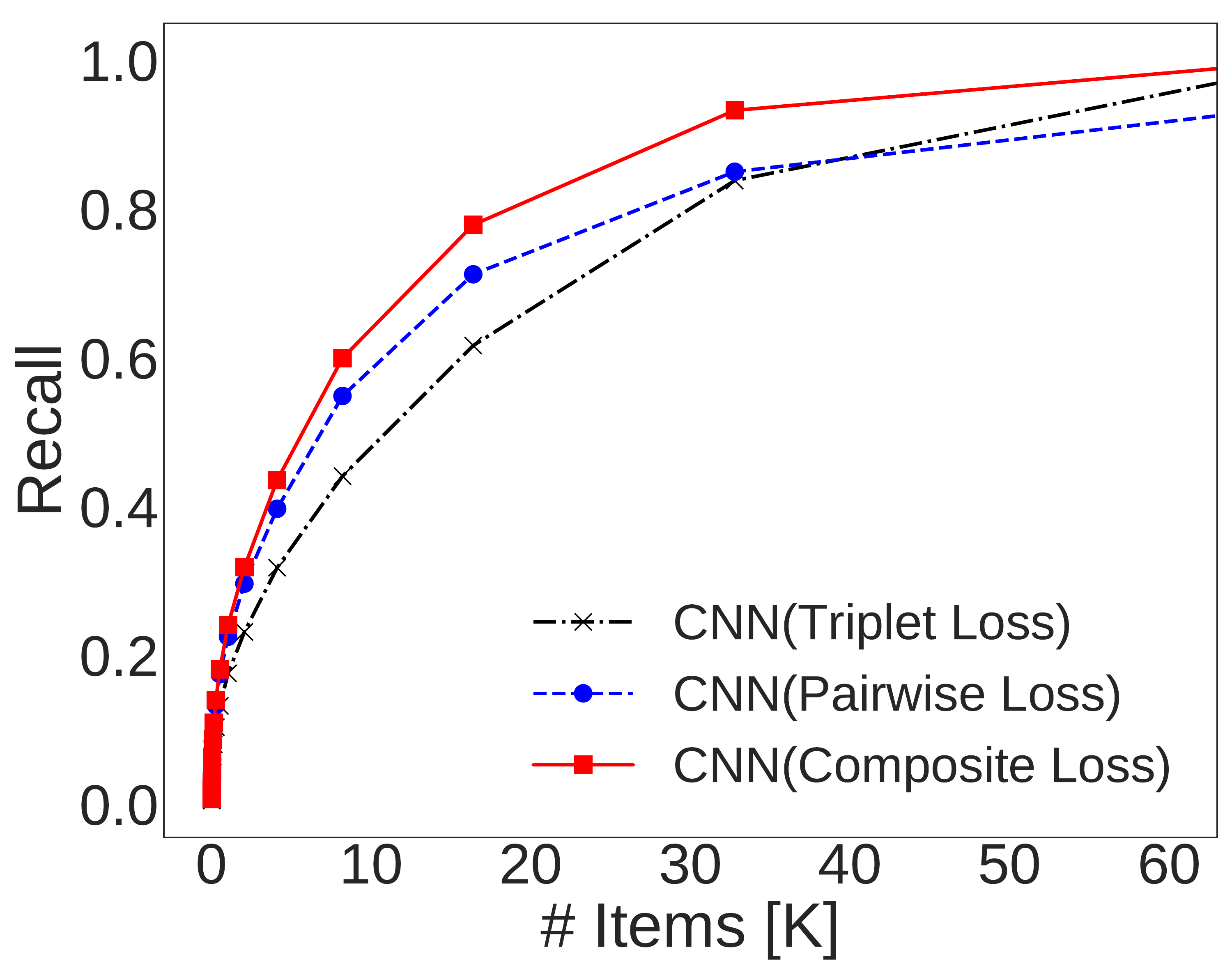} 
		\caption{Loss function}
		\label{fig:loss}
	\end{subfigure}
	\begin{subfigure}{0.48\linewidth}
	\centering 
	\includegraphics[width=\linewidth]{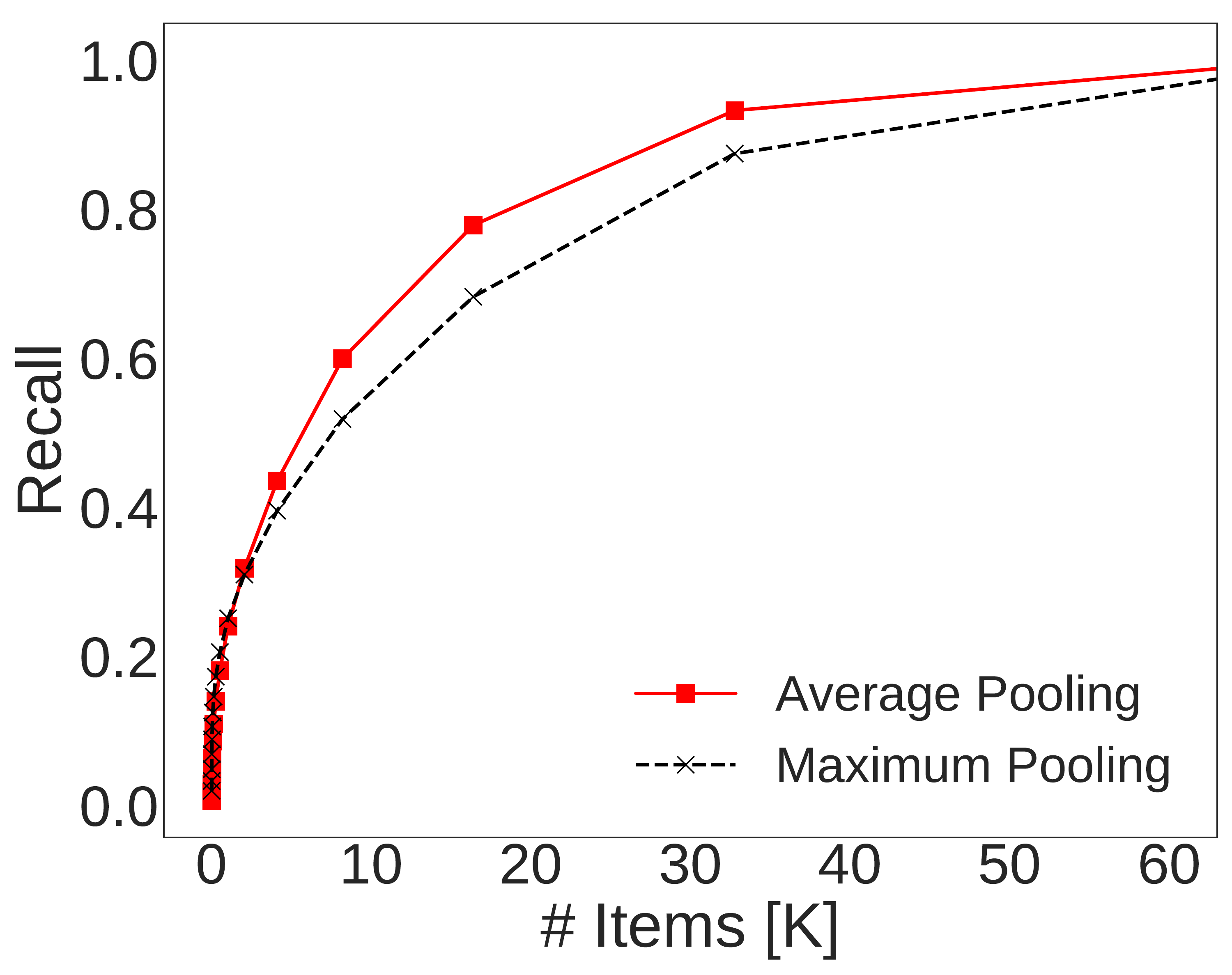} 
	\caption{Pooling function}
	\label{fig:pool}
\end{subfigure}
	\caption{Influence of the hyper-parameters on item-recall performance for Enron dataset (best viewed in color)} 
	\label{fig:ablation}
\end{figure}

\subsection{Influence of Model Parameters}

We evaluate the influence of the hyper-parameters on the performance of CNN embedding in Figure~\ref{fig:ablation}. The dataset is Enron and we use the recall-item curve as the performance measure, for which higher recall means better performance.   

Figure~\ref{fig:epoch} shows that the quality of the embedding improves quickly in the initial stage of training and stabilizes after 50 epochs, which suggests that CNN is easy to train. In Figure~\ref{fig:dim}, we test the performance of CNN using different output dimensions. The results show that the performance improves considerably when increasing the output dimension from 8 to 32 but does not change much afterwards. It suggests that a small output dimension is sufficient for CNN, while CGK and GRU need to use a large output dimension, which slows down distance computation and takes up a large amount of memory.   

Figure~\ref{fig:layer} shows the performance of CNN when increasing the number of convolutional layers from 8 to 12. The results show that the improvements in performance is marginal with more layers and thus there is no need to use a large number of layers. This is favorable as using a large number of layers makes training and inference inefficient. We show the performance of using different number of convolution kernels in a layer in Figure~\ref{fig:channel}. The results show that performance improves when we increase the number of kernels to 8 but drops afterwards.

We report the performance of CNN using different loss functions in Figure~\ref{fig:loss}. Recall that we use a combination of the triplet loss and the approximation error to train CNN. In Figure~\ref{fig:loss}, \textit{Triplet Loss} means using only the triplet loss while \textit{Pairwise Loss} means using only the approximation error. The results show that using a combination of the two loss terms performs better than using a single loss term. The performance of maximum pooling and average pooling is shown in Figure~\ref{fig:pool}. The results show that average pooling performs better than maximum pooling. Therefore, it will be interesting to extend our analysis on maximum pooling in Section~\ref{sec:method} to more pooling methods.

\section{Conclusions}\label{sec:conclusion}

In this paper, we proposed CNN-ED, a model that uses convolutional neural network (CNN) to embed edit distance into Euclidean distance. A complete pipeline (including input preparation, loss function and sampling method) is formulated to train the model end to end and theoretical analysis is conducted to justify choosing CNN as the model structure. Extensive experimental results show that CNN-ED outperforms existing edit distance embedding method in terms of both accuracy and efficiency. Moreover, CNN-ED shows promising performance for edit distance similarity search and is robust to different hyper-parameter configurations. We believe that incorporating CNN embeddings to design efficient string similarity search frameworks is a promising future direction.            

\noindent \textbf{Acknowledgments.} We thank the reviewers for their valuable comments. This work was partially supported GRF 14208318 from the RGC and ITF 6904945 from the ITC of HKSAR.


\bibliographystyle{ACM-Reference-Format}
\balance
\bibliography{string}


\end{document}